\documentclass[acmsmall]{acmart}

\usepackage[capitalise]{cleveref}
\usepackage{paralist}
\usepackage{makecell}
\usepackage{booktabs}
\usepackage{tabularx}
\usepackage{subcaption}
\usepackage{cases}
\usepackage{tikz}

\usetikzlibrary{calc}
\usetikzlibrary{arrows.meta}
\usetikzlibrary{decorations.pathreplacing}
\usetikzlibrary{fit}
\usetikzlibrary{positioning}
\usetikzlibrary{spline}

\newtheorem{insight}{Insight}
\newtheorem{definition}{Definition}
\newtheorem{axiom}{Axiom}

\DeclareMathOperator*{\var}{Var}
\DeclareMathOperator*{\expectation}{\mathbb{E}}

\AtBeginDocument{%
  \providecommand\BibTeX{{%
    \normalfont B\kern-0.5em{\scshape i\kern-0.25em b}\kern-0.8em\TeX}}}

\setcopyright{none}
\copyrightyear{2018}
\acmYear{2018}
\acmDOI{10.1145/1122445.1122456}

\acmConference[PERFORMANCE '21]{PERFORMANCE '21}{November 8--12, 2021}{Milano, Italy}
\acmBooktitle{Woodstock '18: ACM Symposium on Neural Gaze Detection,
  June 03--05, 2018, Woodstock, NY}
\acmPrice{15.00}
\acmISBN{978-1-4503-XXXX-X/18/06}

\renewcommand\footnotetextcopyrightpermission[1]{}

\begin{document}


\title[End-Host Path Selection]{An Axiomatic Perspective on the Performance Effects 
of End-Host Path Selection}


\author{Simon Scherrer}
\author{Markus Legner}
\author{Adrian Perrig}
\affiliation{ETH Zurich, Switzerland}
\author{Stefan Schmid}
\affiliation{TU Berlin, Germany}
\affiliation{University of Vienna, Austria}
\affiliation{Fraunhofer SIT, Germany}

\renewcommand{\shortauthors}{Scherrer et al.}

\begin{abstract}
    In various contexts of networking research, end-host path selection 
    has recently regained momentum as a design principle. While such path selection
    has the potential to increase performance and security of networks,
    there is a prominent concern that it could also lead to network instability
    (i.e., flow-volume oscillation) if paths are selected in a greedy, load-adaptive fashion. 
    However, the extent and the impact vectors of instability caused by path selection
    are rarely concretized or quantified, which is essential 
    to discuss the merits and drawbacks of end-host path selection.
    
    In this work, we investigate the effect
    of end-host path selection on various metrics of networks both qualitatively and quantitatively. 
    To achieve general and fundamental insights,
    we leverage the recently introduced axiomatic perspective
    on congestion control and adapt it to accommodate
    joint algorithms for path selection and congestion control,
    i.e., \emph{multi-path congestion-control} protocols.
    Using this approach, we identify equilibria 
    of the multi-path congestion-control dynamics
    and analytically characterize these equilibria with respect to important metrics of interest in networks (the ``axioms'') such as efficiency, fairness, and loss avoidance.
    Moreover, we analyze how these axiomatic ratings for a general
    network change compared to a scenario without path selection, 
    thereby obtaining an interpretable and quantititative formalization of the
    performance impact of end-host path-selection.
    Finally, we show that there is a fundamental trade-off in multi-path congestion-control protocol
    design between efficiency, stability, and loss avoidance on one side
    and fairness and responsiveness on the other side.
\end{abstract}

%\keywords{}

\maketitle

\section{Introduction}
\label{sec:introduction}

Path selection performed by end-points is a promising
approach to improve efficiency, security, and robustness
of communication networks in their various forms:
To name a few examples, solutions based on end-point path selection 
have been proposed for routing on multiple optimality criteria~\cite{sobrinho2020routing},
multi-tenant data centers~\cite{shahbaz2019elmo},
mobile ad-hoc networks~\cite{johnson1996dynamic}, 
LEO satellite networks~\cite{giuliari2020internet},
intra-domain forwarding~\cite{filsfils2015segment}, 
and inter-domain forwarding~\cite{barrera2017scion}.
% As illustrated by these proposals, end-point path control 
% has become a valuable tool for future networks
% to meet the ever-increasing requirements
% in terms of flexibility, traffic volumes, and threat resistance. 
However, proposals based on end-host
path selection often encounter a
stability concern: researchers have
identified the problem that 
uncoordinated path-selection decisions by end-points
may lead to persistent \emph{oscillation}, i.e.,
an alternating grow-and-shrink pattern of traffic volumes on 
links~\cite{fischer2009adaptive, shaikh2001evaluating}.
The risk of oscillation still represents an obstacle to deployment of path-aware 
networks~\cite{irtf-panrg-what-not-to-do} and
gives rise to schemes that
try to avoid oscillation~\cite{fischer2009adaptive, elwalid2002mate, nelakuditi2002adaptive, kelly2005stability, scherrer2020incentivizing}. While there is a rich literature
presenting solutions for oscillation suppression,
relatively little is known about \emph{how
exactly} and \emph{by how much} 
instability from path selection deteriorates network performance.
In other words, the solution to the oscillation problem
is much clearer than both the impact vectors and the magnitude of
the problem. 

In this work, we therefore aim at qualifying
and quantifying the effects of oscillatory path selection on
various metrics of a network. To tackle this challenge, we must take into account
that end-points in real path-aware networks
employ algorithms which jointly perform path selection and congestion control (CC),
i.e., multi-path congestion-control (MPCC) algorithms.  In this work, we will focus
on MPCC algorithms that are inspired
by greedy, myopic path-selection behavior
and thus simultaneously produce and react to oscillation.
Furthermore, we require an analytical approach that
(i)~captures the congestion-window fluctuations
that represent the oscillation, and 
(ii)~is general enough to deliver fundamental
insights into the nature of CC-assisted
end-host path selection. Alas, 
fluid models~\cite{wardrop1952road, peng2013multipath, key2007path, kelly2005stability, han2006multi}
are well suited to represent equilibria in terms
of the rough traffic distribution on a network;
these models, however, either completely disregard
congestion-control dynamics (such as the classic Wardrop
model~\cite{wardrop1952road}) or
fail to capture the small-scale dynamics
of congestion-window fluctuations 
(as noted by Peng et al., who themselves operate 
with a fluid model~\cite{peng2013multipath}).
More applied approaches, as employed in the
design of multi-path TCP (MPTCP)~\cite{wischik2011design, raiciu2011coupled, khalili2013mptcp},
can capture oscillatory 
phenomena~(e.g., the `flappiness' of protocols~\cite{khalili2013mptcp}), 
but these approaches rely on
ad-hoc reasoning from stylized network examples
and experimental validation, which reduces
their viability as generic analytic tools. 
% As a result, these approaches are better suited
% for the design of specific MPCC protocols
% than for a general analysis of 
% path-selection-induced instability and its
% performance effects.

We argue that a so-called \emph{axiomatic} approach
recently initiated by Zarchy et al.~\cite{zarchy2019axiomatizing}
offers both the right analytical resolution and the
required generality for the question at hand.
This approach is axiomatic in a sense borrowed
from economics and game theory, where
properties with obvious desirability
(e.g., the acyclicity of 
preferences~\cite{blair1982acyclic} or
the fairness of a bargaining 
outcome~\cite{nash1950bargaining})
are formulated as axioms.
Zarchy et al.\ apply this approach to
congestion control by capturing
desirable properties of CC protocols
such as efficiency, fairness, and stability
in axioms. The approach allows to analytically
rate protocols with respect to these axioms and
highlight the fundamental trade-offs between them.
In our work, we further extend Zarchy et
al.'s model to a multi-path
context with the goal of characterizing fundamental
properties of joint algorithms for path selection
and CC.

\subsection{Contribution}
\label{sec:introdcution:contribution}

Our paper uses a theoretical model to investigate how network performance
is affected by the instability due to
greedy end-point path selection. 
In contrast to earlier theoretical
models,
we develop a model that is able to capture both 
path-selection dynamics and
congestion-window fluctuations in \S\ref{sec:model}.
Within this model, we identify and formalize
different classes of dynamic equilibria
(in \S\ref{sec:lossless} and \S\ref{sec:lossy}) 
to which the flow dynamics
can be expected to converge exponentially fast.
These equilibria are essential for the
analytical rating of MPCC protocols:
In \S\ref{sec:axioms}, we rate 
these dynamic equilibria
with respect to a number of performance
metrics (the axioms), which are inspired by 
the recently developed axiomatic
approach to CC~\cite{zarchy2019axiomatizing},
but extended to accommodate path selection. This equilibrium
formalization allows to derive the following
insights in~\S\ref{sec:insights}:
\begin{itemize}
    \item \textbf{No trade-off between efficiency, convergence and
    loss avoidance:}
    Through appropriate protocol tuning, the metrics
    efficiency, loss avoidance,
    and convergence can be simultaneously optimized.
    Hence, there is no trade-off between
    these properties in theory. 
    % Unfortunately, these optimal parameters 
    % depend on system parameters 
    % such as the number of end points and the bottleneck capacities, 
    % making them hard to determine in most practical settings.
    \item \textbf{Trade-off with fairness and responsiveness:}
    There is, however, a fundamental trade-off
    between the above metrics and the fairness and the responsiveness
    of a MPCC protocol. In particular,
    higher responsiveness makes a protocol less efficient,
    but more fair.
    \item \textbf{Effects of introducing end-host path selection:}
    % In addition, we axiomatically characterize the
    % performance of a general network without path selection
    % and compare it to the axiomatic ratings of the MPCC equilibria
    % that arise in the same network given path selection.
    By contrasting the axiomatic performance ratings
    for a general network with and without path selection,
    we obtain a multifaceted formalization of the performance
    impact of introducing end-host path selection.
    This formalization allows to interpret and quantify
    how unstable path selection affects network performance
    depending on network parameters. The insights
    gained from this approach show that there
    are both benefits and drawbacks of
    end-host path selection.
\end{itemize}

\section{Model and Assumptions}
\label{sec:model}

% In this section, we present our model and
% the assumptions that underpin the
% analysis in this work.

\begin{table}
    \centering\small
    \caption{Notation used in our model in alphabetic order.}
    \begin{tabularx}{\textwidth}{rX}
        \toprule
         \textbf{Symbol} & \textbf{Description} \\
         \midrule
         $A=[N]$ & Set of agents in network\\
         $A_{\pi}(t)$ & Set of agents using path~$\pi$ at time~$t$\\
         $a_{\pi}(t)$ & Number of agents using path~$\pi$ at time~$t$\\
         $\alpha(\tau)$ & Additive increase given continuity time~$\tau$\\
         $\beta$ & Multiplicative-decrease parameter\\
         $C$ & Total bottleneck capacity of network\\
         $C_{\pi}$ & Bottleneck capacity of path~$\pi$\\
         $\mathit{cwnd}_i(t)$ & Congestion-window size of agent~$i$ at time~$t$\\
         $f(t)$ & Combined congestion-window size of all agents at time~$t$\\
         $f_{\pi}(t)$ & Combined congestion-window size of all agents using path~$\pi$ at time~$t$\\
         $M_{\pi}(t)$ & Set of agents who migrate away from path~$\pi$ at time~$t$\\
         $m$ & Responsiveness (probability of switching to more attractive path in each time step)\\
         $N$ & Number of agents in the network\\
         $P$ & Number of paths in the network\\
         $\Pi$ & Set of paths in the network\\
         $\pi_i(t)$ & Path used by agent~$i$ at time~$t$\\
         $\pi_{\min}(t)$ & Path with lowest utilization at time~$t$\\
         $r$ & Reset softness (multiplicative decrease of congestion-window size on path switch)\\
         $\mathrm{rank}(\pi, t)$ & Rank of path~$\pi$ at time~$t$ (number of paths with higher utilization than~$\pi$ at time~$t$)\\
         $\tau$ & Continuity time (time since last loss or path switch)\\
         $\tau_i(t)$ & Continuity time of agent~$i$ at time~$t$\\
        %  $u_{\pi}(t)$ & Bottleneck utilization of path~$\pi$ at time~$t$ ($u_{\pi}(t) = f_{\pi}(t)/C_{\pi}$)\\
         $z(a_{\pi}(t),N)$ & Scaling factor for extrapolating
         on-migration flow volume from path flow\\
         \bottomrule
    \end{tabularx}
    \label{tab:notation}
\end{table}

\subsection{Discrete Model}
\label{sec:model:discrete}

\begin{figure}
    \centering
    \begin{tikzpicture}[
    agentnode/.style={circle, draw=black!60, shading=radial,outer color={rgb,255:red,137;green,207;blue,240},inner color=white, thick, inner sep=0pt, minimum size=6mm},
    cwndsquare/.style={draw=black!60, shading=radial,outer color={rgb,255:red,137;green,207;blue,240},inner color=white, thick, minimum size=6mm},
    pathsquare/.style={draw=black!60, shading=radial,outer color={rgb,255:red,240;green,240;blue,240},inner color=white, thick},
]

    \definecolor{ag0}{rgb}{0.54,0.81,0.94};
    \definecolor{ag1}{rgb}{0.53,0.66,0.42};
    \definecolor{ag2}{rgb}{1.0,0.6,0.4};
    \definecolor{agN}{rgb}{0.98, 0.91, 0.71}
    \definecolor{bostonuniversityred}{rgb}{0.8, 0.0, 0.0}

    % ------------------------------------------------
    % Time t
    \node[minimum width=53mm] at (1.9, 5.35) {\textbf{Time} $\boldsymbol{t}$};
    
    % Agents
    \node[align=center] at (-1.7, 4.9) {\textbf{Agent}\\\textbf{set}~$\boldsymbol{A}$\textbf{,}\\$\boldsymbol{|A| = N}$};
    \node[agentnode]                    at (-1.7, 3.5) {$0$};
    \node[agentnode,outer color=ag1]    at (-1.7, 2.5) {$1$};
    \node[agentnode,outer color=ag2]    at (-1.7, 1.5) {$2$};
    \node                               at (-1.7, 0.5) {...};
    \node[agentnode,outer color=agN]    at (-1.7, -0.5) {$\overline{N}$};
    
    % ~~~~~~~~~~~~~ Paths
    
    % Path 0
    \node[pathsquare,minimum width=53.5mm,minimum height=22mm] at (1.9, 3.96) (t1p0) {};
    \node[below right = 0.5mm and 0.5mm of t1p0.north west] (t1p0label) {\textbf{Path} $\boldsymbol{\pi}$};
    \node[below right = 5mm and 0.5mm of t1p0.north west] (t1p0a) {$A_{\pi}(t) = \{$};
    \node[agentnode,outer color=ag1, right = -1mm of t1p0a] (t1p0a1) {$1$};
    \node[right = -0.5mm of t1p0a1] {,};
    \node[agentnode,outer color=agN,right= 2mm of t1p0a1] (t1p0aN) {$\overline{N}$};
    \node[right=-0.5mm of t1p0aN] {, ... $\}$};
    \node[cwndsquare,outer color=ag1,minimum width=25mm,below right = 14mm and 1mm of t1p0.north west] (t1p0cwnd1) {$\mathit{cwnd}_1(t)$};
    \node[cwndsquare,outer color=agN,minimum width=22mm,right = -0.4mm of t1p0cwnd1] (t1p0cwndN) {$\mathit{cwnd}_{\overline{N}}(t)$};
    \node[cwndsquare,outer color=white,minimum width=5mm,right = -0.4mm of t1p0cwndN] (t1p0rest) {...};
    \draw [decorate,decoration={brace,amplitude=5pt,aspect=0.85}] (t1p0cwnd1.north west) -- (t1p0rest.north east) node [black,midway,yshift=0.6cm,xshift=2cm] {$f_{\pi}(t)$}; 
    
    % Path 1
    \node[pathsquare,minimum width=53.5mm,minimum height=22mm] at (1.9, 1.85) (t1p1) {};
    \node[below right = 0.5mm and 0.5mm of t1p1.north west] (t1p1label) {\textbf{Path} $\boldsymbol{\pi'}$};
    \node[below right = 5mm and 0.5mm of t1p1.north west] (t1p1a) {$A_{\pi'}(t) = \{$};
    \node[agentnode,outer color=ag0, right = -1mm of t1p1a] (t1p1a0) {$0$};
    \node[right=-0.5mm of t1p1a0] {, ... $\}$};
    \node[cwndsquare,outer color=ag0,below right = 14mm and 1mm of t1p1.north west] (t1p1cwnd0) {$\mathit{cwnd}_0(t)$};
    \node[cwndsquare,outer color=white,minimum width=30mm,right = -0.4mm of t1p1cwnd0] (t1p1rest) {...};
    \draw [decorate,decoration={brace,amplitude=5pt,aspect=0.85}] (t1p1cwnd0.north west) -- (t1p1rest.north east) node [black,midway,yshift=0.6cm,xshift=1.6cm] {$f_{\pi'}(t)$}; 
    
    \node[pathsquare,minimum width=53.5mm,minimum height=5mm] at (1.9, 0.5) (t1pSpare) {...};
    
    % Path 2
    \node[pathsquare,minimum width=53.5mm,minimum height=22mm] at (1.9, -0.8) (t1p2) {};
    \node[below right = 0.5mm and 0.5mm of t1p2.north west] (t1p2label) {\textbf{Path} $\boldsymbol{\pi''}$};
    \node[below right = 5mm and 0.5mm of t1p2.north west] (t1p2a) {$A_{\pi''}(t) = \{$};
    \node[agentnode,outer color=ag2, right = -1mm of t1p2a] (t1p2a2) {$2$};
    \node[right=-0.5mm of t1p2a2] (t1p2bracket) {, ... $\}$};
    \node[cwndsquare,outer color=ag2,minimum width=20mm,below right = 14mm and 1mm of t1p2.north west] (t1p2cwnd2) {$\mathit{cwnd}_2(t)$};
    \node[cwndsquare,outer color=white,right = -0.4mm of t1p2cwnd2] (t1p2rest) {...};
    \draw [decorate,decoration={brace,amplitude=5pt,aspect=0.85}] (t1p2cwnd2.north west) -- coordinate [above right=2mm and 9mm] (t1p2tip) (t1p2rest.north east) node {};
    \node[right=1mm of t1p2bracket] (t1p2flabel) {$f_{\pi''}(t)$};
    \draw[-latex] (t1p2tip) -- (t1p2flabel);
    
    % ------------------------------------------------
    % Time t + 1
    \node[minimum width=53mm] at (7.4, 5.35) (t2) {\textbf{Time} $\boldsymbol{t+1}$};
    
    % Path 0
    \node[pathsquare,minimum width=53.5mm,minimum height=22mm] at (7.4, 3.96) (t2p0) {};
    \node[below right = 0.5mm and 0.5mm of t2p0.north west] (t2p0label) {\textbf{Path} $\boldsymbol{\pi}$};
    \node[below right = 5mm and 0.5mm of t2p0.north west] (t2p0a) {$A_{\pi}(t+1) = \{$};
    \node[agentnode,outer color=agN, right = -1mm of t2p0a] (t2p0aN) {$\overline{N}$};
    \node[right=-0.5mm of t2p0aN] (t2p0bracket) {, ... $\}$};
    \node[cwndsquare,outer color=agN,minimum width=22mm,below right = 14mm and 1mm of t2p0.north west] (t2p0cwndN) {$\mathit{cwnd}_{\overline{N}}(t)$};
    \node[cwndsquare,outer color=agN,minimum width=3mm,right = -0.4mm of t2p0cwndN] (t2p0cwndN2) {};
    \node[cwndsquare,outer color=white,minimum width=5mm,right = -0.4mm of t2p0cwndN2] (t2p0rest) {...};
    \draw [decorate,decoration={brace,amplitude=5pt,aspect=0.85}] (t2p0cwndN.north west) -- coordinate [above right=2mm and 10.5mm] (t2p0tip) (t2p0rest.north east) node {};
    \node[right=1mm of t2p0bracket] (t2p0flabel) {$f_{\pi}(t+1)$};
    \draw[-latex] (t2p0tip) -- (t2p0flabel);
    
    \node[pathsquare,minimum width=53.5mm,minimum height=5mm] at (7.4, 0.5) (t2pSpare) {...};
    
    % Path 1
    \node[pathsquare,minimum width=53.5mm,minimum height=22mm] at (7.4, 1.85) (t2p1) {};
    \node[below right = 0.5mm and 0.5mm of t2p1.north west] (t2p1label) {\textbf{Path} $\boldsymbol{\pi'}$};
    \node[below right = 5mm and 0.5mm of t2p1.north west] (t2p1a) {$A_{\pi'}(t+1) = \{$};
    \node[agentnode,outer color=ag0, right = -1mm of t2p1a] (t2p1a0) {$0$};
    \node[right=-0.5mm of t2p1a0]  (t2p1bracket) {, ... $\}$};
    \node[cwndsquare,outer color=ag0,below right = 14mm and 1mm of t2p1.north west] (t2p1cwnd0) {$\mathit{cwnd}_0(t)$};
    \node[cwndsquare,outer color=ag0,minimum width=3mm,right = -0.4mm of t2p1cwnd0] (t2p1cwnd02) {};
    \node[cwndsquare,outer color=white,minimum width=20mm,right = -0.4mm of t2p1cwnd02] (t2p1rest) {...};
    \draw [decorate,decoration={brace,amplitude=5pt,aspect=0.85}] (t2p1cwnd0.north west) -- coordinate [above right=2mm and 13.5mm] (t2p1tip) (t2p1rest.north east) node  {};
    \node[right=1mm of t2p1bracket] (t2p1flabel) {$f_{\pi'}(t+1)$};
    \draw[-latex] (t2p1tip) -- (t2p1flabel);
    \draw [decorate,decoration={brace,amplitude=5pt,mirror}] (t2p1cwnd0.south west) -- (t2p1cwnd02.south east) node [black,midway,yshift=-0.4cm] {$\mathit{cwnd}_0(t+1)$};
    
    % Path 2
    \node[pathsquare,minimum width=53.5mm,minimum height=22mm] at (7.4, -0.8) (t2p2) {};
    \node[below right = 0.5mm and 0.5mm of t2p2.north west] (t2p2label) {\textbf{Path} $\boldsymbol{\pi''}$};
    \node[below right = 5mm and 0.5mm of t2p2.north west] (t2p2a) {$A_{\pi''}(t+1) = \{$};
    \node[agentnode,outer color=ag1, right = -1mm of t2p2a] (t2p2a1) {$1$};
    \node[right=-0.5mm of t2p2a1] {,};
    \node[agentnode,outer color=ag2, right = 2mm of t2p2a1] (t2p2a2) {$2$};
    \node[right=-0.5mm of t2p2a2] (t2p2bracket) {, ... $\}$};
    \node[cwndsquare,outer color=ag1,minimum width=15mm,below right = 14mm and 1mm of t2p2.north west] (t2p2cwnd1) {$\mathit{cwnd}_1(t+1)$};
    \draw [decorate,decoration={brace,amplitude=5pt,mirror}] (t2p2cwnd1.south west) -- (t2p2cwnd1.south east) node [black,midway,yshift=-0.6cm,align=center] {$\mathit{cwnd}_1(t+1)$\\$= r\cdot \mathit{cwnd}_1(t)$};
    \node[cwndsquare,outer color=ag2,minimum width=20mm,right = -0.4mm of t2p2cwnd1] (t2p2cwnd2) {$\mathit{cwnd}_2(t)$};
    \node[cwndsquare,outer color=ag2,minimum width=3mm,right = -0.4mm of t2p2cwnd2] (t2p2cwnd22) {};
    \node[cwndsquare,outer color=white,minimum width=8mm,right = -0.4mm of t2p2cwnd22] (t2p2rest) {...};
    \draw [decorate,decoration={brace,amplitude=5pt,aspect=0.85}] (t2p2cwnd1.north west) --  coordinate [above right=2mm and 18mm] (t2p2tip) (t2p2rest.north east) node {};
    \node[above right=-1mm and -8mm of t2p2bracket] (t2p2flabel) {$f_{\pi''}(t+1)$};
    \draw[-latex] (t2p2tip) -- (t2p2flabel);

    \draw [decorate,decoration={brace,amplitude=5pt}] (t2p0.north east) -- (t2p2.south east) node [black,midway,xshift=0.8cm,align=center] {\textbf{Path}\\\textbf{set} $\boldsymbol{\Pi}$,\\
    $\boldsymbol{|\Pi| = P}$};

    % ------------------------------------------------
    % Time t + 1
    \node[minimum width=53mm] at (10.8, 5.25) {\textbf{...}};
    
    % Migration
    \draw[-latex,bostonuniversityred,dotted,ultra thick] (t1p0a1.south) to[spline through={(t1p0cwnd1.east)(4,2)(4.5,-0.5)}] (t2p2a1.south west);

\end{tikzpicture}
    \vspace*{-15pt}
    \caption{Illustration of discrete model (Notation: $\overline{N} = N-1$).  The dotted arrow visualizes path migration by agent~1 from
    path~$\pi$ to path~$\pi''$.}
    \label{fig:model:illustration}
\end{figure}
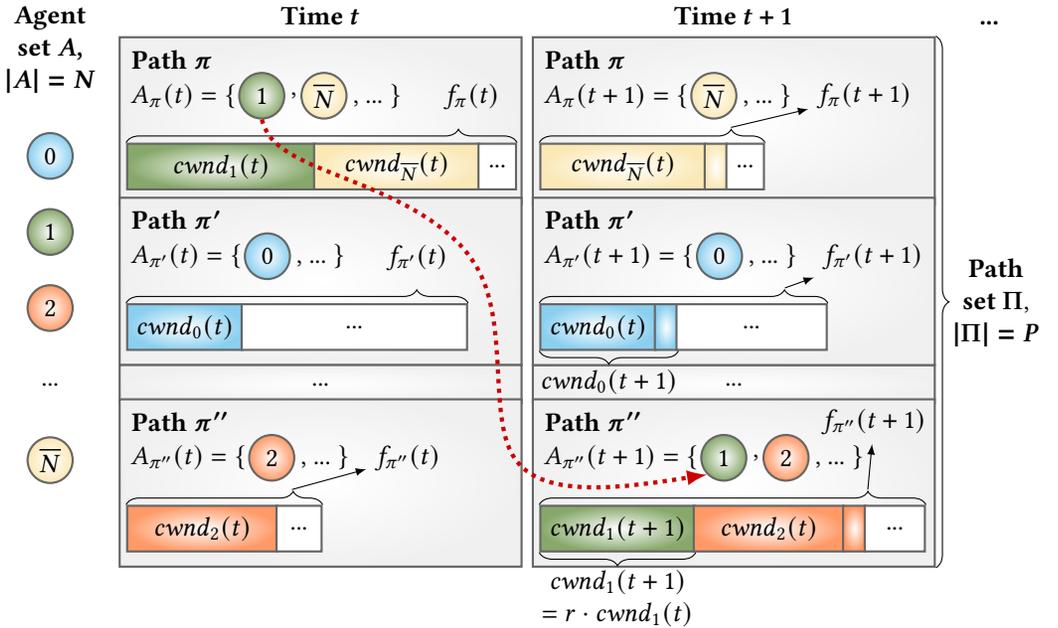

We leverage the analytical model of congestion control
proposed by Zarchy et al.~\cite{zarchy2019axiomatizing}
and extend it to a multi-path context with path selection
as illustrated in~\cref{fig:model:illustration}. 
In summary, $N$ agents (denoted by set~$A = [N] := \{0,...,N-1\}$) 
compete for bandwidth on the bottleneck links of~$P$ 
parallel paths from set~$\Pi$. Each agent~$i \in A$ maintains
a congestion window with size~$\mathit{cwnd}_i$, which evolves over time~$t$. 
At each moment~$t \in \mathbb{N}_0$ in discrete
time, any path~$\pi\in\Pi$ accommodates a set~$A_\pi(t)$ of agents
that use path~$\pi$ at moment~$t$, and carries 
load~$f_{\pi}(t) = \sum_{i \in A_{\pi}(t)} \mathit{cwnd}_i(t)$.
Moreover, in each time step~$t$, every agent~$i$ takes two actions.
First, agent~$i$ performs \emph{congestion control}, i.e., adapts
its congestion-window size~$\mathit{cwnd}_i(t)$ according
to a chosen CC protocol~$CC_i(t,\mathit{cwnd}_i(t))$,
resulting in congestion-window size~$\mathit{cwnd}_i(t+1)$.
Second, agent~$i$ performs \emph{path selection}, i.e.,
determines the path~$\pi$ such that~$i \in A_{\pi}(t)$
and~$\mathit{cwnd}_i(t)$ is included in~$f_{\pi}(t)$,
according to a given path-selection strategy.
In~\cref{fig:model:illustration} as well as 
in our following analysis, agents
implement probabilistic greedy path selection,
i.e., switch to the path carrying the lowest load
in the last time step with a given probability~$m$.
Finally, in order to investigate different behaviors
for congestion-window adaptation upon path switches,
we introduce a reset-softness parameter~$r \in [0,1]$ 
that determines the extent of congestion-window reduction 
for path-switching agents (e.g., agent~1 
in~\cref{fig:model:illustration}).

The agents are further constrained by path capacities~$C_\pi$, $\pi \in \Pi$,
where~$C_{\pi}$ is the amount of
data in maximum segment size (MSS) 
that can be transmitted on path~$\pi$ 
during one round-trip time (RTT).
If the capacity~$C_{\pi}$ of path~$\pi$ is 
exceeded by the flow~$f_{\pi}(t)$, the agents~$A_{\pi}(t)$
experience packet loss and take this loss into account
in their congestion-control protocol.\footnote{We 
note that this loss modelling
is a simplification in three respects.
First, loss may already occur 
when~$f_\pi(t) > s_{\pi}$, namely if all agents send
out all traffic~$f(t)$ in a burst that exceeds
the buffer size~$s_{\pi}$. Second, even
if~$f_\pi(t) > C_\pi$, the loss may not be
perceived by all agents. Third,
CC algorithms may react differently depending on the number of recent losses.}
For example, the TCP Reno protocol, with a multiplicative decrease
of~0.5 as a reaction to loss and an additive increase
of~1 otherwise,
is modelled as follows for an agent~$i$ using path~$\pi$
at time~$t$:
\begin{equation}
    \mathit{TCPReno}(t, \mathit{cwnd}_i(t)) = \begin{cases}
        \mathit{cwnd}_i(t) + 1 & \text{if } f_\pi(t) \leq C_\pi\\
        0.5 \cdot \mathit{cwnd}_i(t) & \text{otherwise}
    \end{cases}
\end{equation}

% In summary, the single-path CC dynamics $f(t)$ 
% in the model of Zarchy et al.~\cite{zarchy2019axiomatizing} are
% uniquely determined by the number~$N$ of competing agents,
% a bottleneck capacity~$C$, 
% an initial choice $\{\mathit{cwnd}_i(0)\}_{i\in [N]}$ 
% of congestion-window size for all agents, 
% and a choice~$\{\mathit{CC}_i(t)\}_{i \in [N]}$
% of the CC protocol employed by each agent.

\subsection{Scenario of Interest and Assumptions}
\label{sec:model:scenario}

Since the goal of this work is to characterize the 
worst-case effects of oscillatory path selection,
our analysis throughout the paper will focus
on a network scenario that maximizes the severity
of load oscillation. This scenario
has the following properties, which henceforth
serve as assumptions:

\textbf{Greedy load-adaptive path selection.}
Oscillation is caused by
greedy, myopic path selection
behavior~\cite{scherrer2020incentivizing}, 
which dynamically determines 
the number~$a_{\pi}(t) = |A_\pi(t)|$ 
of agents on path~$\pi$.
In any time step~$t$, agents seek out 
the path~$\pi_{\min}(t)$ with the lowest 
bottleneck utilization~$f_{\pi}(t)/C_{\pi}$ 
and hence the lowest latency 
(assuming roughly equal propagation delay of all paths
as stated below)
and lowest loss rate.
Since monitoring
the state of alternative paths and switching paths consume resources, 
agents may not consider a path change 
in every time step. Instead, the path-selection behavior is regulated by a 
\emph{path-migration probability}~$m \in (0,1]$,
denoting the probability with which an agent switches to
a more attractive path in any time step.
Alternatively, $m$ can be interpreted as a measure
for the \emph{responsiveness} of agents.

\textbf{Sequential multi-path usage.} The intensity of
oscillations grows with the size of shifted flow volume
per time unit. In order to maximize oscillation,
we therefore assume that a path-switching agent 
completely stops using its previously used path and
exclusively sends on the newly selected path. This
coarse-granular migration behavior produces sequential
instead of concurrent usage of multiple paths.
This mode of sequential
multi-path usage approximates the actual behavior
of real-world algorithms such as MPTCP, 
which tends to use only the most attractive path
for data transmission and sends
a negligible amount of probing traffic
over the alternative paths~\cite{wischik2011design, khalili2013mptcp, kelly2005stability}.
% Moreover, simultaneous usage of multiple paths
% may lead to severe performance reduction
% due to packet reordering,
% which remains a challenge to 
% date~\cite{alheid2014analysis}. 
Moreover, the average utility improvement
per user that is possible by concurrently
using multiple paths instead of a single
selected path vanishes for a high number
of agents~\cite{wang2011cost}.
Sequential multi-path usage implies that
$\sum_{\pi\in\Pi} a_{\pi}(t )= N\ \forall t$.

\textbf{Disjoint and similar paths.} We investigate 
a network consisting of paths that are parallel, disjoint 
and equal in terms of latency~$D_{\pi}$ and 
bottleneck capacity~$C_{\pi} = C/P$, where~$C$
is the bottleneck capacity of the complete network. 
Such a network, while being a simplification of
general networks, is likely to bring
out the worst-case effects of myopic, greedy
path selection, which are the subject of this paper.
In particular, load oscillations are strongest if the
actions of the sending agents
are strongly correlated because they react
to the same (potentially misleading) 
feedback signals (i.e., path loss and latency) 
simultaneously~\cite{scherrer2020incentivizing}. 
If agents sharing a link
react to different feedback
signals or at different times, 
e.g., because they are using different
paths with different round-trip latencies, 
their actions are less strongly
correlated and the
flow dynamics are likely to oscillate less.
The feedback synchronization by equal path RTTs also
ensures that the discrete time steps of the model 
have consistent duration across all paths.

\subsection{Stochastic Dynamics}
\label{sec:model:stochastic-dynamics}

In summary, a multi-path congestion-control
protocol~$\mathit{MPCC}(\mathit{CC}, m, r)$ is 
a combination of a CC protocol~$\mathit{CC}(t)$,
a responsiveness parameter~$m$, and a reset-softness 
parameter~$r$. In a network with path selection,
the MPCC dynamics can thus be represented by a 
pair of functions $(a_{\pi}(t), f_{\pi}(t))$
for any path~$\pi \in \Pi$.  
Since the path-selection behavior is probabilistic
(regulated by responsiveness parameter~$m$),
the MPCC dynamics are not uniquely determined by initial conditions,
but need to be modeled as a stochastic process. In particular,
the MPCC dynamics under universal adoption of 
$\mathit{MPCC}(\mathit{CC}, m, r)$,
are given by \begin{subequations}%
\allowdisplaybreaks%
\begin{align}
    a_{\pi}(t+1) &= 
    \begin{cases}
        a_{\pi}(t) - |M_{\pi}(t)| & \text{if } \pi \neq \pi_{\min}(t)\\
        a_{\pi}(t) + \sum_{\tilde\pi\neq\pi} |M_{\tilde\pi}(t)| & \text{otherwise}
    \end{cases}\label{eq:agents:raw}\\
    f_{\pi}(t+1) &=
    \begin{cases}
         f_{\pi}(t) - \sum_{i \in M_{\pi}(t)} \mathit{cwnd}_i(t) + \sum_{i \in A_{\pi}(t)\setminus M_{\pi}(t)} \Delta\mathit{cwnd}_i(t) &  \text{if } \pi \neq \pi_{\min}(t)\\ 
         f_{\pi}(t) + \sum_{\tilde\pi \neq \pi}\sum_{j \in M_{\tilde\pi}(t)} r\cdot\mathit{cwnd}_j(t) + \sum_{i \in A_{\pi}(t)} \Delta\mathit{cwnd}_i(t) & \text{otherwise},
    \end{cases}\label{eq:flow:raw}
\end{align}\label{eq:raw}%
\end{subequations}%
where~$M_{\pi}(t)$ is a random subset of~$A_{\pi}(t)$, 
which contains the agents who leave path~$\pi$
at time~$t$, and 
$\Delta\mathit{cwnd}_i(t)=\mathit{cwnd}_i(t+1)-\mathit{cwnd}_i(t)$.
Intuitively, the flow on a more congested path~$\pi$ 
is reduced by the congestion windows of all 
agents~$M_{\pi}$ that leave the path, 
and increased by the congestion-window growth 
of the remaining 
agents~$A_{\pi}(t)\setminus M_{\pi}(t)$.
In contrast, the flow on the least congested path~$\pi$ is 
increased by the reset congestion-window 
sizes~$r\cdot\mathit{cwnd}_j(t)$ of the 
agents~$j \in M_{\tilde\pi}$ 
who migrate to path~$\pi$ and the congestion-window 
growth of the previously present agents~$A_{\pi}(t)$.

\subsection{Expected Dynamics}
\label{sec:model:expected}

While the formulations in~\cref{eq:raw} 
capture the evolutionary dynamics of an MPCC system,
their discrete and probabilistic nature hinders 
analytic treatment. However, as we investigate
large-scale systems with a high number of agents,
the law of large numbers allows that the probabilistic
elements in~\cref{eq:raw} can be well approximated
by their expected values and traffic randomness
can be greatly ignored. For the remainder of this paper,
we therefore consider the \emph{expected} MPCC dynamics,
where the recursion on the 
random variables~$\big(a_\pi(t), f_\pi(t)\big)$ 
is approximated with a recursion on the 
expectations~$\big(\hat{a}_\pi(t), \hat{f}_\pi(t)\big)$
(where we write~$\hat{x}:=\mathbb{E}[x]$ for any function~$x$).
The accuracy of this approximation will be
validated with simulations in~\cref{sec:appendix:approximation-accuracy}.

Concerning the agent dynamics in~\cref{eq:agents:raw}, 
we note that~$\mathbb{E}[|M_{\pi}(t)|] = m\cdot \hat{a}_{\pi}(t)$
for any path~$\pi \neq \pi_{\min}(t)$. 
Moreover, the expected volume of flow associated
with the agents in~$M_{\pi}$ in~\cref{eq:flow:raw} is a proportional
share of the expected total flow~$\hat{f}_{\pi}(t)$ on path~$\pi$:
$\mathbb{E}[\sum_{i \in M_{\pi}(t)}\mathit{cwnd}_i(t)] = m\cdot \hat{f}_{\pi}(t)$.
By the same argument, it holds that
$\mathbb{E}[\sum_{\tilde\pi \neq \pi}\sum_{j \in M_{\tilde\pi}(t)} r\cdot\mathit{cwnd}_j(t)] = m\cdot r \cdot \sum_{\tilde\pi\neq\pi} \hat{f}_{\tilde\pi}(t)$
for~$\pi = \pi_{\min}(t)$. However, in order
to make the second case of~\cref{eq:flow:raw}
independent of flows $f_{\tilde\pi}$ on alternative paths,
we additionally make the following approximation: 
$\sum_{\tilde\pi\neq\pi} \hat{f}_{\tilde\pi}(t) \approx \big(N-\hat{a}_{\pi}(t)\big)/\hat{a}_{\pi}(t) \cdot \hat{f}_{\pi}(t) = z(\hat{a}_{\pi}(t), N)\cdot \hat{f}_{\pi}(t)$, 
where~$z(\hat{a}_{\pi}(t),N)$ is
henceforth referred to as the \emph{extrapolation 
factor}.
In this approximation, the flow on path~$\pi$ is scaled
proportionally to the number of 
agents~$N-\hat{a}_{\pi}(t)$ on other paths.
This approximation can be justified on the grounds that
in a steady state, 
imbalances in path load are likely to stem from imbalances
in the number of agents between paths, not from
imbalances in the average congestion-window size between paths.

Finally, in order to arrive at the expected flow dynamics
$\hat{f}_{\pi}(t)$, the expected combined congestion-window
change~$\mathbb{E}[\sum_{k\in A_{\pi}(t)}\Delta\mathit{cwnd}_k(t)]$ 
(or for~$A_{\pi}(t+1)$, respectively) must be formalized.
Of course, this change depends on the CC protocols employed
by the agents. In order to maximize the generality of
our analysis, we rely on the following generic form 
of a loss-based CC protocol employed by each agent~$i$, where 
$\pi_{i}(t)$ denotes the path that agent $i$ uses at time $t$:
\begin{equation}
\mathit{CC}_i(t,\ \mathit{cwnd}_i(t)) =
\begin{cases}
\mathit{cwnd}_i(t) + \alpha\big(\tau_{i}(t)\big) & \text{if } f_{\pi_i(t)}(t) \leq C_{\pi_i(t)}\\
\beta \cdot \mathit{cwnd}_i(t) & \text{otherwise}
\end{cases}
\label{eq:cc-protocol}
\end{equation}
Here,~$\tau_i(t)$ is the so-called \emph{continuity time}
of agent~$i$, i.e., the number of time steps in which agent~$i$
has already been on its current path without experiencing packet loss.
This continuity time is the argument to a function~$\alpha$,
which determines the additive increase to the congestion window
in absence of loss.
This formulation allows to mimic the window-growth
behavior in classic TCP Reno~\cite{mo1999analysis},
in the widely deployed TCP CUBIC~\cite{ha2008cubic},
in the slow-start phase of many TCP protocols~\cite{stevens1997tcp}, 
or in more theoretical MIMD protocols~\cite{altman2005fairness}.
%+The values of~$\alpha$ can be expected to remain below 3~\cite{xu2016congestion}.
Finally, $\beta \in [0,1]$ is a parameter that determines the
multiplicative decrease of the congestion-window size in the
case of packet loss, which is the predominant practice
in CC protocols. 
% Clearly, this protocol allows to represent
% only loss-based CC protocols; similar to 
% Zarchy et al.~\cite{zarchy2019axiomatizing}, we will analyze
% latency-based CC protocols by means of a specific axiom.

Based on the probability distribution for 
the continuity time~$\tau_i(t)$
of any agent~$i \in A_{\pi}(t)$ at time~$t$ 
from~\cref{sec:equilibria:distribution}, 
we can calculate the average congestion-window increase
per agent conditioned on the path~$\pi$ used by the agent at time~$t$: 
$\hat{\alpha}_\pi(t) = \sum_{\tau = 0}^{\infty} \mathbb{P}\left[\tau_{i}(t) = \tau\ \middle|\ i \in A_{\pi}(t)\right] \cdot \alpha(\tau)$.
This average congestion-window increase
then allows to obtain the aggregate additive
increase in absence of loss. In contrast,
loss reduces the expected flow volume~$\hat{f}_{\pi}(t)$
through multiplicative decrease~$\beta$,
complementing the effects of out-migration 
(for~$\pi \neq \pi_{\min}(t)$) or 
in-migration (for~$\pi_{\min}(t)$).
Under universal adoption
of a protocol~$\mathit{MPCC}(\mathit{CC},m,r)$,
% the expectation of the aggregate congestion-window
% change of non-migrating agents is thus given by
% \begin{equation}
%     \mathbb{E}\Big[\sum_{i\in A_{\pi}(t)\setminus M_{\pi}(t)}\Delta\mathit{cwnd}_i(t)\Big] = 
%     \begin{cases}
%         (a_{\pi}(t) - \mathbb{E}\big[|M_{\pi}(t)|\big]) \cdot \hat{\alpha}_\pi(t) & \text{if } f_{\pi}(t) \leq C_{\pi}\\
%         -(1-\beta) \cdot \sum_{i\in A_{\pi}(t)\setminus M_{\pi}(t)} \mathit{cwnd}_i(t) & \text{otherwise}
%     \end{cases}
% % \end{equation}
%
% Hence, the expected dynamics are as follows:
the expected dynamics therefore are:
\begin{subequations}\allowdisplaybreaks%
\begin{align}
    \hat{a}_{\pi}(t+1) &= \begin{cases}
        (1-m)\cdot \hat{a}_{\pi}(t) & \text{if } \pi \neq \pi_{\min}(t)\\
        (1-m)\cdot \hat{a}_{\pi}(t) + m\cdot N & \text{otherwise}
        \label{eq:agents:expected}
    \end{cases}\\
    \hat{f}_{\pi}(t+1) &= \begin{cases}
        (1-m)\cdot \hat f_{\pi}(t) + \hat\alpha_\pi(t)\cdot(1-m)\cdot \hat{a}_{\pi}(t) & \text{if } \pi \neq \pi_{\min}(t) \land \hat{f}_{\pi}(t) \leq C_{\pi}\\ 
        \big(1+m\cdot r\cdot z(\hat{a}_{\pi}(t),N)\big)\cdot \hat{f}_{\pi}(t) + \hat\alpha_{\pi}(t)\cdot \hat{a}_{\pi}(t) &  \text{if } \pi = \pi_{\min}(t) \land \hat{f}_{\pi}(t) \leq C_{\pi}\\
        \beta \cdot (1-m) \cdot \hat{f}_{\pi}(t) & \text{if } \pi \neq \pi_{\min}(t) \land \hat{f}_{\pi}(t) > C_{\pi}\\
        \big(\beta+m\cdot r\cdot z(\hat{a}_{\pi}(t),N)\big)\cdot \hat{f}_{\pi}(t) & \text{if } \pi = \pi_{\min}(t) \land \hat{f}_{\pi}(t) > C_{\pi}
    \label{eq:flow:expected}
    \end{cases}
\end{align}\label{eq:expected}
\end{subequations}

\subsection{Limitations}
\label{sec:model:limitations}

While our model presents a tractable approach to analyze 
oscillatory MPCC dynamics, our investigation and the resulting 
insights have clear limitations worth addressing in
future research. In particular, as our network model is an extension
of the network model by Zarchy et al.~\cite{zarchy2019axiomatizing},
our work inherits some limitations noted by Zarchy et al., most importantly
the assumption of synchronized feedback, the focus on a specific
type of network, and the disregard for queuing dynamics.
However, it is noteworthy that our work addressed the 
previously identified challenge concerning randomized protocols 
through the concept of expected dynamics. In general,
the comprehensiveness of our analysis would benefit
from relaxing the worst-case conditions elicited
in~\cref{sec:model:scenario}, most prominently 
the assumption of disjoint and similar paths, and from
introducing latency-based and model-based CC protocols.
\section{Lossless Equilibria}
\label{sec:lossless}

% \begin{table}
%     \centering
%     \caption{Notation introduced in~\cref{sec:lossless}.}
%     \begin{tabular}{rl}
%         \toprule
%          \textbf{Symbol} & \textbf{Description} \\
%          \midrule
%         $\hat{a}^{(p)}$ & Rank-$p$ equilibrium value for expected agent dynamics\\
%         $\hat{a}_{\pi}^{(p)}(t)$ & Rank-$p$ trajectory function for expected agent dynamics on path~$\pi$\\
%         $\hat{f}^{(p)}$ & Rank-$p$ hypothetical equilibrium value for expected flow dynamics\\
%         $\hat{f}_{\pi}^{(p)}(t)$ & Rank-$p$ trajectory function for expected flow dynamics on path~$\pi$\\
%         $\hat{f}^{(p)\ast}$ & Initial equilibrium flow volume for rank-$p$ path in lossy equilibrium\\
%         \bottomrule
%     \end{tabular}
%     \label{tab:notation:equilibria}
% \end{table}

In order to rate MPCC protocols, we focus on
the \emph{equilibria} that these protocols induce,
i.e., stable load patterns to which the MPCC dynamics 
from~\cref{eq:expected} eventually converge.
In this section, we characterize one class of
equilibria  that are attained before the capacity limit
of any bottleneck link is exceeded, i.e.,
these equilibria are \emph{lossless}. 
Equilibria without this lossless property, i.e.,
lossy equilibria, are presented
in~\cref{sec:lossy}. All of 
these equilibria are \emph{dynamic}
equilibria, i.e.,
periodic patterns of the number of
agents and the load on the different paths.
Note that the insights regarding equlibria only apply 
to the theoretical construct
of expected dynamics in an exact sense, 
and only approximately
apply to actual MPCC dynamics.

\subsection{Structure of Lossless Equilibria}
\label{sec:lossless:structure}

In order to characterize lossless equilibria,
we need to investigate whether the expected MPCC 
dynamics tend to exhibit a certain pattern
in the case where capacity limits
are disregarded. Unfortunately, even this simplified
discrete dynamical system
(determined by~\cref{eq:expected} without the two
last cases of~\cref{eq:flow:expected}) is
analytically intractable due to the presence of
case distinctions in the evolution 
functions~\cite{galor2007discrete}.
Instead, we use a hybrid approach, similar to previous 
work~\cite{akella2002selfish}: By performing simulations
as in~\cref{fig:approximation-accuracy},
we arrive at the following two observations
about MPCC dynamics with greedy, myopic
agents sharing parallel and similar paths (cf.~\cref{sec:model:scenario}), 
which serve  as a basis for further analytical investigation:

\textbf{In-migration is utilization-maximizing:}
    Whenever path~$\pi$ 
    with minimal utilization within the expected dynamics, i.e., $\hat{u}_{\pi}(t) = \hat{f}_{\pi}(t)/(C/P)$,
    experiences in-migration according to the second
    case of~\cref{eq:flow:expected}, this path tends to 
    become the most utilized path in the next time step.\footnote{This
    observation suggests that
    myopic, greedy load-adaptive
    path selection is not a
    Nash equilibrium strategy,
    which has also been demonstrated
    by recent 
    research~\cite{scherrer2020incentivizing}.}
    
    \textbf{Out-migration is order-preserving:} 
    If two paths~$\pi$ and~$\tilde\pi$ 
    with~$\hat{u}_{\pi}(t) > \hat{u}_{\tilde\pi}(t)$
    experience out-migration according to the first
    case of~\cref{eq:flow:expected}, it tends
    to hold that~$\hat{u}_{\pi}(t+1) > \hat{u}_{\tilde\pi}(t+1)$.
    
If the expected dynamics consistently conform to these
two observations, they exhibit the following pattern
which uniquely determines the least utilized path
in every time step:
\begin{definition}
MPCC dynamics exhibit \textbf{\textit{$\boldsymbol{P}$-step oscillation}}
if there exists a time~$t_0 \geq 0$ such that
\begin{equation}
\forall T \geq 0.\quad\mathrm{rank}(\pi, t_0) = p \implies \mathrm{rank}(\pi, t_0 + T) = (p + T) \bmod P,
\end{equation}
% where $\pi^{(p)}(t)$ is the \textbf{\textit{path with rank~$\boldsymbol{p}$}}, 
% i.e., the path with  the $(p+1)$-largest utilization at time~$t$:
% \begin{equation}\pi^{(p)}(t) = \pi \quad \text{s.t.} \quad |\{\tilde\pi\ |\ \hat{u}_{\tilde\pi}(t) > \hat{u}_{\pi}(t)\}| = p.\end{equation}
where $\mathrm{rank}(\pi, t)$ ranks 
all paths~$\pi \in \Pi$ in descending order according to 
their utilization at time~$t$:
\begin{equation}\mathrm{rank}(\pi, t) = p \iff |\{\tilde\pi\ |\ \hat{u}_{\tilde\pi}(t) > \hat{u}_{\pi}(t)\}| = p.\end{equation}
\label{def:p-step}
\vspace{-20pt}
\end{definition}

In $P$-step oscillation, the assignment of the rank to paths
changes in a round-robin fashion, i.e., in any time step~$t$,
every path~$\pi$ rises by one rank,
except the path with rank~$P-1$
(i.e., with the lowest expected utilization), 
which obtains rank 0 at time~$t+1$. 
% For example, it holds $\mathrm{rank}(\pi,t) = 0 \implies \mathrm{rank}(\pi,t+1) = 1$,
% whereas $\mathrm{rank}(\pi,t) = P-1 \implies \mathrm{rank}(\pi,t+1) = 0$.
After~$P$ time steps, a path reaches its original
place in the ranking order, i.e., $\mathrm{rank}(\pi,t) = \mathrm{rank}(\pi,t+P)$
for all~$t\geq t_0$.
We present an argument for
the prevalence of $P$-step oscillation 
in~\cref{sec:lossless:flow}.

\subsection{Lossless Agent Equilibrium}
\label{sec:lossless:agents}

As this $P$-step oscillation
uniquely determines the
least congested path in any time step $t \geq t_0$,
this pattern also determines the agent-migration dynamics.
Starting from an agent distribution 
$\{a_{\pi}(t_0)\}_{\pi\in\Pi}$ at time~$t_0$,
all the paths~$\pi$ with $\mathrm{rank}(\pi,t_0) \neq P-1$ will
experience an outflow of agents (according
to case 1 in~\cref{eq:agents:expected}) and
only the path with rank $P-1$ experiences an
inflow of agents (according to case 2
in~\cref{eq:agents:expected}). In a single round
of $P$-step oscillation with start time~$t_0$, 
the path~$\pi^{(0)}$ with $\mathrm{rank}(\pi^{(0)}, t_0) = 0$ 
will thus first experience agent 
outflow for~$P-1$ times
and then once experience agent inflow.
Hence, the following difference equation characterizes the discrete dynamical
system for a granularity of $P$ time steps:
\begin{equation}
    \hat{a}_{\pi^{(0)}}(t_0 + P) = (1-m)^P\cdot \hat{a}_{\pi^{(0)}}(t_0) + m\cdot N,
    \label{eq:agent-dynamics:diff-equation}
\end{equation}

To find an equilibrium of the dynamic system for
the agent dynamics on~$\pi^{(0)}$,
we identify a fixed point of the difference equation in~\cref{eq:agent-dynamics:diff-equation},
i.e., we solve 
\begin{equation}\hat{a}^{(0)} = (1-m)^P\cdot \hat{a}^{(0)} + m\cdot N \quad \iff \quad \hat{a}^{(0)} = \frac{m\cdot N}{1-(1-m)^P},\end{equation}
where $\hat{a}^{(0)}$ is the \emph{equilibrium value}
for any~$\hat{a}_{\pi}(t)$ with $\mathrm{rank}(\pi,t_0) = 0$,
which generalizes as follows:

\begin{insight}
    \textbf{Convergence to Unique Dynamic Agent Equilibrium.} Under $P$-step oscillation, the expected agent dynamics $\{\hat{a}_{\pi}(t)\}_{\pi\in\Pi}$ of
    an MPCC system asymptotically converge to a unique dynamic equilibrium,
    i.e., a cyclic series of states. 
    This dynamic equilibrium of the agent dynamics
    consists of $P$~states in each of which
    the rank-$p$ path accommodates the corresponding
    equilibrium amount of agents~$\hat{a}^{(p)}$,
    i.e., 
    \begin{equation}
    \hat{a}_{\pi}(t) = \hat{a}^{(\mathrm{rank}(\pi,t))}, \text{ where }
    \hat{a}^{(p)} = \frac{(1-m)^p\cdot m \cdot N}{1-(1-m)^P}.
    \end{equation}
    \label{ins:agent-equilibrium}
    \vspace{-10pt}
\end{insight}

\begin{figure}
\centering
\begin{minipage}{.48\textwidth}
  \centering
  \includegraphics[width=\linewidth,trim=0 10 10 0]{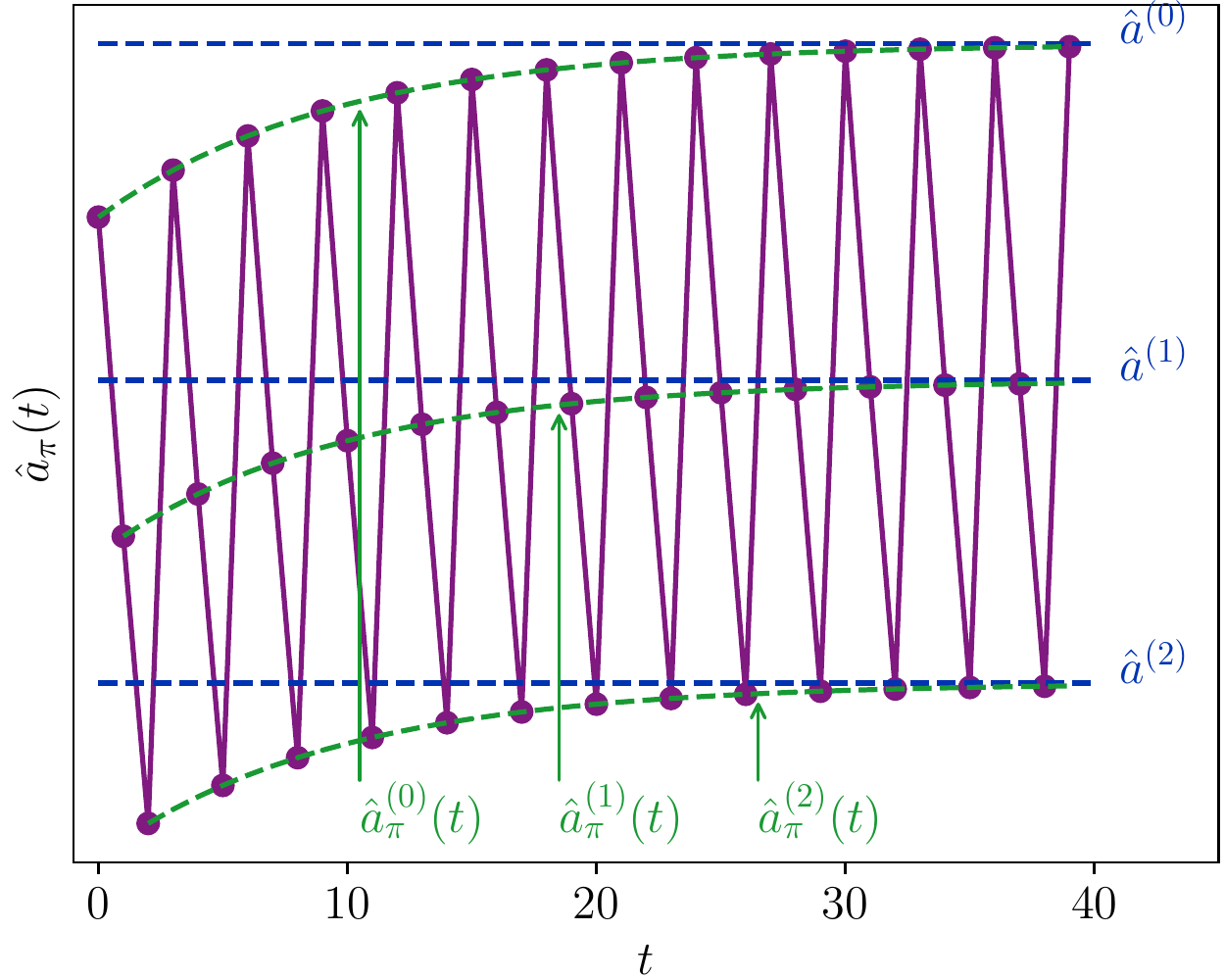}
  \caption{Convergence to lossless agent equilibrium
  for~$P=3$ and $m=0.1$.}
  \label{fig:equilibrium:agent}
\end{minipage}%
\hfill
\begin{minipage}{.48\textwidth}
  \centering
  \includegraphics[width=\linewidth,trim=10 10 0 0]{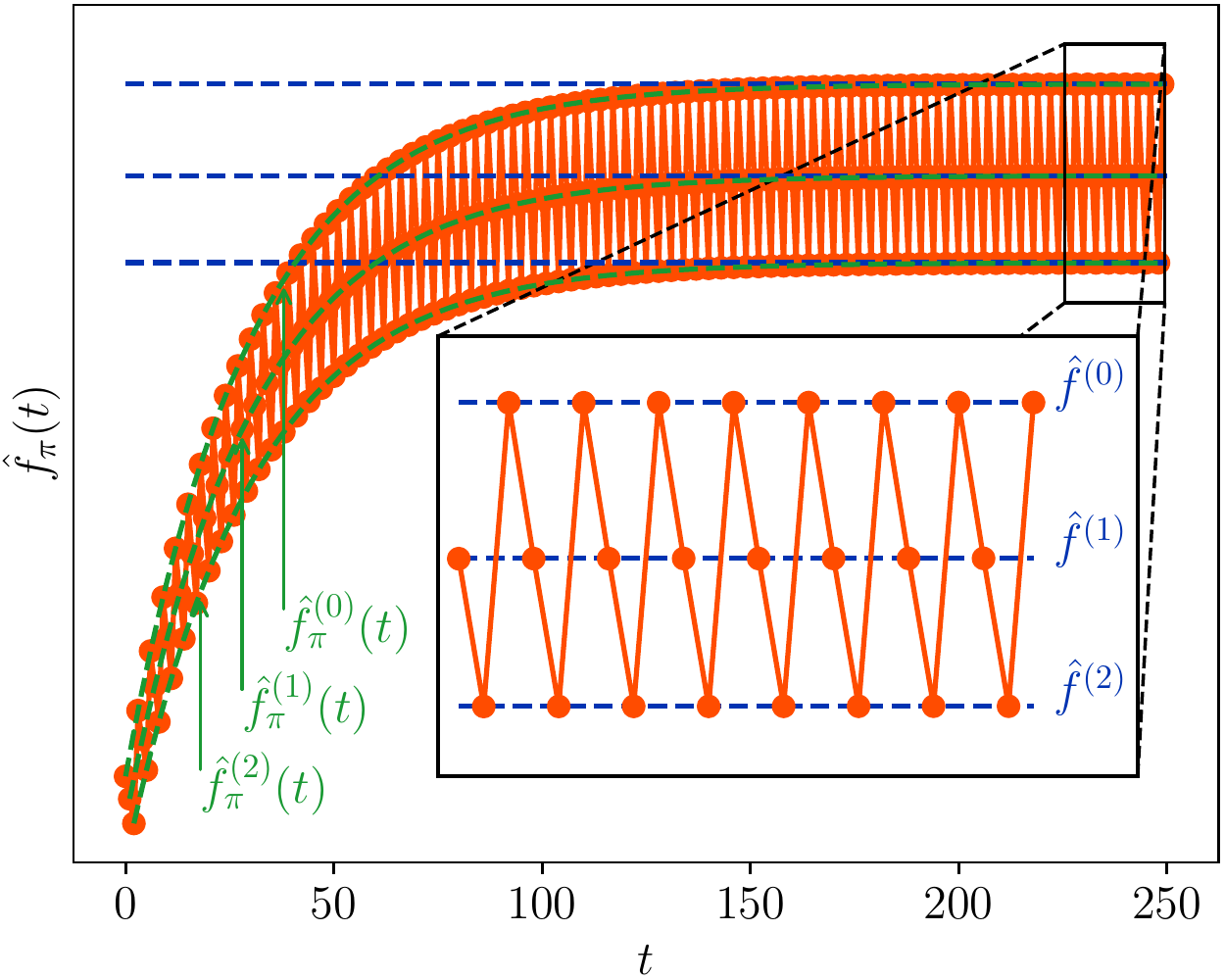}
  \caption{Convergence to lossless
  flow equilibrium for $P=3$, $m=0.1$, and~$r=0.5$
  .}
  \label{fig:equilibrium:hypothetical}
\end{minipage}%
\end{figure}

This convergence can be shown 
by finding a \emph{trajectory function}:
\begin{definition}
    A \textbf{trajectory function~$x_{\pi}^{(p)}(t)$}
    is an explicit interpolation 
    function that yields the 
    correct value of path-specific dynamics~$x_{\pi}(t)$
    at all moments where path~$\pi$ has rank~$p$:
    \begin{equation}
        \forall k \in \mathbb{N}_{\geq0}. \quad  x_{\pi}^{(p)}(t_{\pi p}+k\cdot P) = x_{\pi}(t_{\pi p}+k\cdot P),
    \end{equation}
    where $t_{\pi p} = \min \{t\ |\ t\geq t_0 \land \mathrm{rank}(\pi, t) = p\}$ and~$t_0$ is
    the start time of $P$-step oscillation.
    \label{def:trajectory}
\end{definition}

For the agent dynamics~$\hat{a}_{\pi}(t)$, 
such a trajectory function is given by
\begin{equation}
    {\hat{a}}^{(p)}_{\pi}(t) =  \left(\hat{a}_{\pi}(t_{\pi p}) - \hat{a}^{(p)}\right) \cdot (1-m)^{t-t_{\pi p}} + \hat{a}^{(p)}.
\end{equation} 
As $\lim_{t\rightarrow\infty}  \hat{a}^{(p)}_{\pi}(t) = \hat{a}^{(p)}$,
the trajectory functions converge to the 
equilibrium found above exponentially fast.
\Cref{fig:equilibrium:agent}
visualizes the asymptotic convergence
to the dynamic 
equilibrium~$\{\hat{a}^{(p)}\}_{p \in [P]}$
(highlighted in blue)
along the trajectory functions.

\subsection{Lossless Flow Equilibrium}
\label{sec:lossless:flow}

After identifying the
agent equilibrium in~\cref{sec:lossless:agents}, 
we identify the equilibria of the MPCC flow 
dynamics~$\{\hat{f}_{\pi}(t)\}_{\pi\in\Pi}$ in
this section.
We first consider \emph{hypothetical
equilibria}, which are equilibria of
the flow dynamics under the assumption
that the capacity of each path 
is never exceeded. In a second step,
we will show under which conditions
these hypothetical equilibria are actual
equilibria.

\subsubsection{Hypothetical Flow Equilibria}
\label{sec:lossless:flow:hypothetical}

To find the hypothetical equilibria
of the flow dynamics, we can simplify
the flow dynamics from~\cref{eq:flow:expected}
by disregarding the capacity limit~$C/P$.
In addition, we insert the equilibrium
agent levels~$\hat{a}^{(p)}$ from~\cref{sec:lossless:agents} 
and the expected additive increase~$\hat\alpha^{(p)}$ 
derived in~\cref{sec:equilibria:distribution} to
arrive at the following formulation:
\begin{equation}
    \hat{f}_{\pi}(t+1) = \begin{cases}
        (1-m)\cdot \left( \hat f_{\pi}(t) + \hat\alpha^{(\mathrm{rank}(\pi, t))}\cdot \hat{a}^{(\mathrm{rank}(\pi, t))}\right) & \text{if } \mathrm{rank}(\pi, t) \neq P-1\\ 
        \big(1+m\cdot r\cdot z(m,P)\big)\cdot \hat{f}_{\pi}(t) + \hat\alpha^{(P-1)} \cdot \hat{a}^{(P-1)} &  \text{if } \mathrm{rank}(\pi, t) = P-1,
    \label{eq:flow:hypothetical}
    \end{cases}
\end{equation}
where the extrapolation
factor~$z$ is only dependent on~$m$ and~$P$
given the agent equilibrium, i.e.,
$z(m, P) = N/\hat{a}^{(P-1)} -1 = (1-(1-m)^{P-1})/(m\cdot(1-m)^{P-1})$.

Similar to~\cref{eq:agent-dynamics:diff-equation},
we set up a first-order difference
equation for the dynamics for
the path that has rank~$p$ at time~$t_0$
(where the $P$-step oscillation starts)
and find a fixed point that is attained
every~$P$ time steps, for example
for ranks~0 and~$P-1$:
% \begin{equation}
%     \hat{f}^{(0)} = (1+m\cdot r \cdot z)\cdot(1-m)^{P-1}\cdot \hat{f}^{(0)} +
%     \big((1+m\cdot r \cdot z)\cdot (\sum_{p = 0}^{P-2} \hat{\alpha}^{(p)}) + \hat\alpha^{(P-1)}\big) \cdot \hat{a}^{(P-1)}
% \end{equation}
\begin{subequations}%
\begin{align}
    \hat{f}^{(0)} &= \frac{\big((1+m\cdot r \cdot z(m,P))\cdot (\sum_{p = 0}^{P-2} \hat{\alpha}^{(p)}) + \hat\alpha^{(P-1)}\big) \cdot \hat{a}^{(P-1)}}{1-(1+m\cdot r \cdot z(m,P))\cdot(1-m)^{P-1}},
    \label{eq:flow:fixed-point:rank-0}\\
    \hat{f}^{(P-1)} &= \frac{\big(\sum_{p = 0}^{P-2} \hat{\alpha}^{(p)} + \hat\alpha^{(P-1)}\cdot(1-m)^{P-1}\big) \cdot \hat{a}^{(P-1)}}{1-(1+m\cdot r \cdot z(m,P))\cdot(1-m)^{P-1}}.
    \label{eq:flow:fixed-point:rank-P-1}
\end{align}%
\end{subequations}%
The fixed point for a general rank~$p$ can
be derived analogously and expressed by
a similar (albeit quite complicated) term
$\hat{f}^{(p)}$ shown in~\cref{eq:lossless:rank-p}
in~\cref{sec:appendix:p-step-oscillation-consistency}.
These fixed points~$\{\hat{f}^{(p)}\}_{p\in[P]}$
constitute the hypothetical equilibrium,
i.e., if a rank-$p$ path carries flow 
volume~$\hat{f}^{(p)}$, the path will
carry this flow volume again~$P$ time steps
later, where it is again the rank-$p$ path.

\begin{insight}
    \textbf{Hypothetical Dynamic Flow Equilibrium.} 
    If capacity limits of links are disregarded,
    the dynamic equilibrium of 
    the flow dynamics~$\{\hat{f}_{\pi}(t)\}_{\pi\in\Pi}$
    consists of~$P$~states in each of which
    the rank-$p$ path accommodates flow volume~$\hat{f}^{(p)}$.
    \label{ins:flow:hypothetical}
\end{insight}

In order for such an equilibrium to be valid, 
it must be consistent
with $P$-step oscillation, i.e., it must hold 
that~$\hat{f}^{(p)} > \hat{f}^{(p+1)}$ for
all~$p\in[P-1]$. Interestingly, 
if a certain parameter combination
is associated with an invalid equilibrium, 
it follows that $P$-step oscillation is fundamentally
impossible for that parameter combination.
However, we show 
in~\cref{sec:appendix:p-step-oscillation-consistency} 
that only a small part of the parameter space,
containing rather extreme parameters,
is inconsistent with~$P$-step oscillation.

Similarly as in~\cref{sec:lossless:agents},
convergence to this equilibrium can be proven
using a trajectory function (cf.~\cref{def:trajectory}). 
The following trajectory function yields the correct flow volume
in all subsequent time steps 
where path~$\pi$ has rank~$p$ again:
\begin{equation}
    \hat{f}^{(p)}_{\pi}(t) = 
    \big(\hat{f}_{\pi}(t_{\pi p}) - \hat{f}^{(p)}\big) \cdot
    \big((1+m\cdot r\cdot z(m,P))\cdot(1-m)^{P-1}\big)^{\frac{t-t_{\pi p}}{P}} 
    + \hat{f}^{(p)}.
    \label{eq:flow:trajectory}
\end{equation}
The limit of this trajectory function
for $t \rightarrow \infty$ is the
equilibrium value~$\hat{f}^{(p)}$,
which establishes convergence;\footnote{
Note that $\hat{f}^{(0)}$ from~\cref{eq:flow:fixed-point:rank-0}
is undefined for~$r = 1$, as the flow dynamics
do not converge to a fixed point in that case.
Given~$r=1$, the trajectory function for rank 0
can be expressed with the following linear function,
which has no limit:
\begin{equation}\hat{f}_{\pi}^{(0)}(t) = \left[(1-m)^{1-P} \cdot \left(\sum\nolimits_{p=0}^{P-2} \hat{\alpha}^{(p)}\right) + \hat{\alpha}^{(P-1)}\right]\cdot\hat{a}^{(P-1)}\cdot P^{-1} \cdot (t-t_{\pi p}) + \hat{f}_{\pi}(t_{\pi0}).\label{eq:flow:trajectory:r1}\end{equation}}
% The asymptotic convergence
% to the hypothetical 
% equilibrium~$\{\hat{f}^{(p)}\}_{p \in [P]}$
this is illustrated 
in~\cref{fig:equilibrium:hypothetical}.

\subsubsection{Actual Flow Equilibrium}
\label{sec:lossless:flow:actual}

Intuitively, this hypothetical equilibrium
given by~$\{\hat{f}^{(p)}\}_{p\in[P]}$
is an actual equilibrium of the MPCC
dynamics if the convergence is not
disturbed by the capacity limit~$C/P$ on any path~$\pi$,
i.e., if the trajectory functions for all ranks
consistently remain below~$C/P$.
We therefore require an upper bound
on all trajectory 
functions~$\{\hat{f}^{(p)}_{\pi}(t)\}_{p\in [P]}$.
Thanks to the structure of $P$-step oscillation,
it holds that~$\hat{f}^{(p)} > \hat{f}^{(p+1)}$
$\forall p \in [P-1]$. Therefore,
in the hypothetical equilibrium,
$\hat{f}^{(0)}$ represents an upper bound
on the flow dynamics. We speak
of flow dynamics with \emph{consistent trajectories}
if such an ordering not only holds
on the equilibrium values~$\hat{f}^{(p)}$,
but also on the trajectory 
functions~$\hat{f}_{\pi}^{(p)}(t)$ for all paths~$\pi$:
\begin{definition}
    Flow dynamics~$\{\hat{f}_{\pi}(t)\}_{t\geq0}$ have
    \textbf{consistent trajectories} at time point~$t'$
    if on every path~$\pi \in \Pi$,
    the rank-specific trajectory 
    functions~$\{\hat{f}^{(p)}_{\pi}(t)\}_{p \in [P]}$
    satisfy the following condition:
    \begin{equation}
        \forall p \in [P-1],\ t \geq t'.\quad \hat{f}^{(p)}_{\pi}(t) > \hat{f}^{(p+1)}_{\pi}(t)
    \end{equation}
\end{definition}
% As trajectories are consistent in equilibrium,
% there is always a time point~$t'$ after which
% the trajectory functions are consistent.

As trajectories are always eventually consistent, the
trajectory function~$\hat{f}^{(0)}_{\pi}(t)$
for rank 0 is therefore an upper bound on all
trajectory functions~$\{\hat{f}^{(p)}_{\pi}(t)\}_{p\in [P]}$
and by consequence also an upper bound on the flow dynamics~$\{\hat{f}_{\pi}\}_{t \geq 0}$
for any path~$\pi$.
As~$\hat{f}^{(0)}_{\pi}(t)$ is monotonic,
its function values will not exceed~$C/P$
if~$\hat{f}_{\pi}(t_{\pi p}) \leq C/P$ 
and~$\hat{f}^{(0)} \leq C/P$.
Due to the introduction of capacity limits, 
it is necessary to alter the definition 
of~$t_{\pi p}$ to be the first
point in time after oscillation began (at~$t_0$)
where $\mathrm{rank}(\pi, t_{\pi p}) = p$ (as before) and additionally
$\hat{f}_{\pi}(t_{\pi p}) \leq C/P$.\footnote{We assume that such a~$t_{\pi p}$
always exists as any reasonable CC's reaction to loss
reduces~$\hat{f}_{\pi}(t)$ below~$C/P$
eventually.}
Therefore, we arrive at the following insight:
\begin{insight}
    \textbf{Dynamic Lossless Flow Equilibrium.}
    The hypothetical equilibrium (disregarding capacity limitations)
    from~\cref{ins:flow:hypothetical} is an actual, lossless equilibrium
    (taking capacity limits into account) for the 
    flow dynamics~$\{\hat{f}_{\pi}(t)\}_{\pi\in\Pi}$ 
    if and only if~$\hat{f}^{(0)} \leq C/P$, i.e.,
    the maximum flow-equilibrium level does not
    exceed the bottleneck capacity of any path.
\end{insight}

\section{Lossy Equilibria}
\label{sec:lossy}

In this section, we characterize~\emph{lossy} equilibria,
i.e., dynamic equilibria where ~$\hat{f}^{(0)} > C/P$ and 
the flow dynamics therefore periodically
exceed bottleneck capacities.

\subsection{Structure of Lossy Equilibria}
\label{sec:lossy:structure}

\begin{figure}
\centering
\begin{minipage}{.48\textwidth}
  \centering
  \includegraphics[width=\linewidth,trim=0 0 10 0]{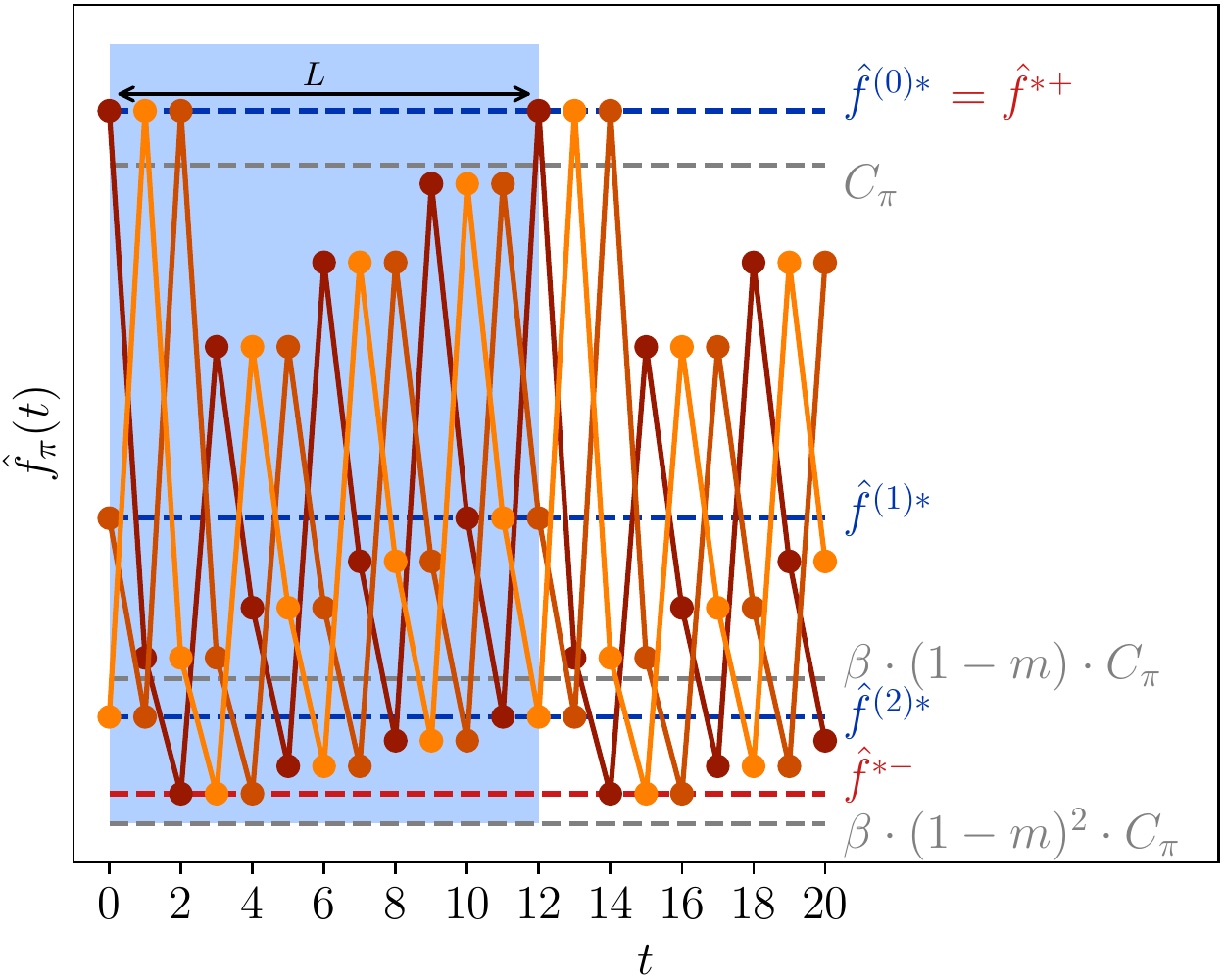}
  \caption{Type-1 lossy equilibrium
  for $P=3$, $m=0.45$, and~$r=0.9$ (One period is highlighted
  in light-blue).}
  \label{fig:equilibrium:loss:type-1}
\end{minipage}
\hfill
\begin{minipage}{.48\textwidth}
  \centering
  \includegraphics[width=\linewidth,trim=10 0 0 0]{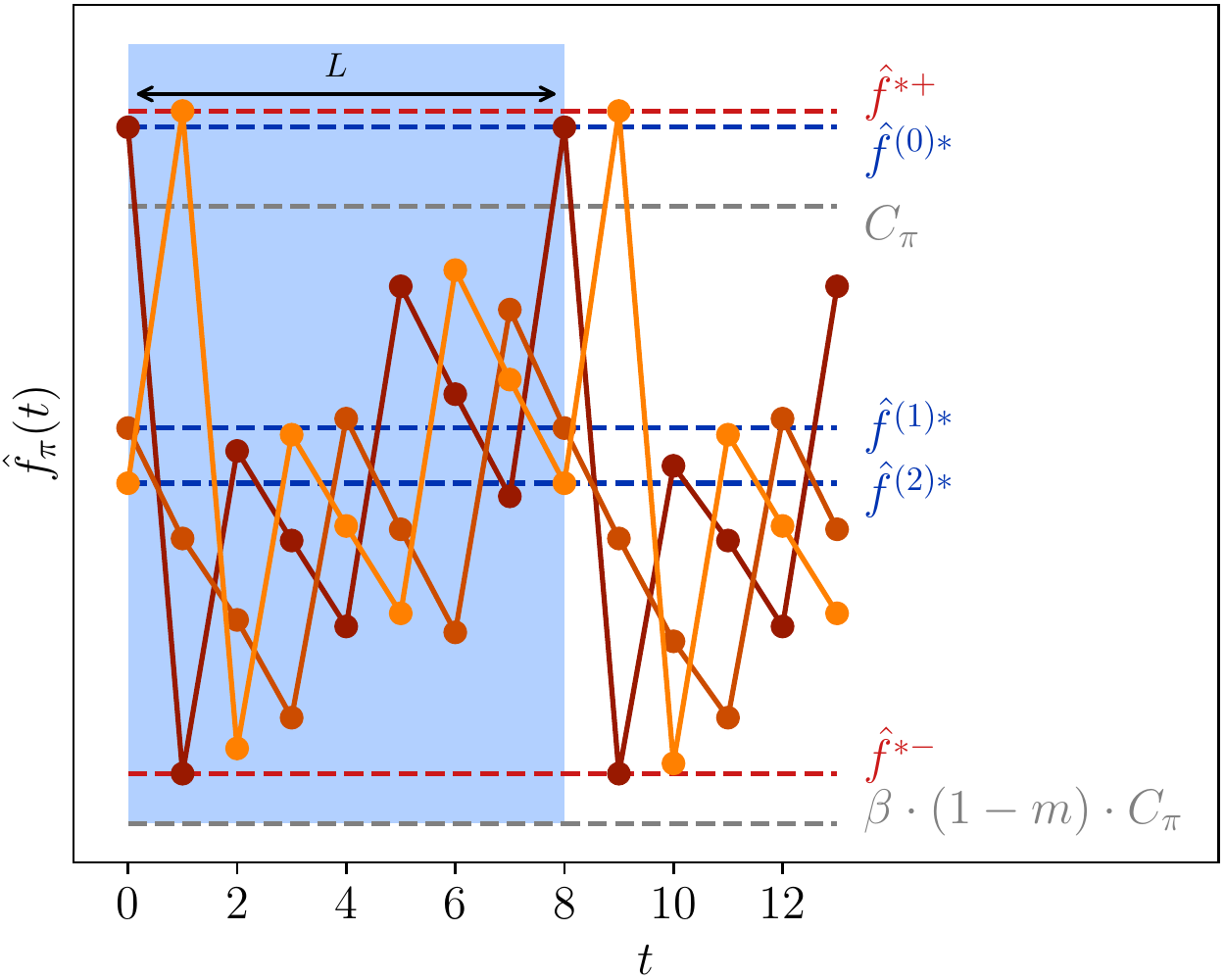}
  \caption{Type-2 lossy equilibrium
  for $P=3$, $m=0.1$, and~$r=1$ (One period is highlighted
  in light-blue).}
  \label{fig:equilibrium:loss:type-2}
\end{minipage}
\end{figure}

In order to identify the typical structure
of lossy equilibria, we again rely on simulations
similar to~\cref{sec:lossless:structure}.
Based on these simulations, we can distinguish
two types of lossy equilibria, illustrated
in~\cref{fig:equilibrium:loss:type-1,fig:equilibrium:loss:type-2}.
Note that both of these lossy-equilibrium
types are characterized by flow 
volumes~$\{\hat{f}^{(p)\ast}\}_{p\in[P]}$,
each carried by the path with rank~$p$
in the state that is designated as the
initial state of the lossy equilibrium
($t=0$ in the figures) and is periodically
revisited every~$L$ time steps. Moreover,
the boundary points, i.e., the largest and 
smallest flow volume
arising in a lossy equilibrium, are
denoted by~$\hat{f}^{\ast+}$ and~$\hat{f}^{\ast-}$,
respectively.

The main distinguishing property of type-1
lossy equilibria (cf.~\cref{fig:equilibrium:loss:type-1})
is that these lossy equilibria are consistent
with~$P$-step oscillation despite the 
occasional multiplicative decrease~$\beta$
on rank-0 paths. In contrast, type-2 lossy
equilibria (cf.~\cref{fig:equilibrium:loss:type-2})
temporarily deviate from~$P$-step oscillation
whenever there is packet loss on a path.
In that case, the rank-0 path with loss
directly becomes the rank-$(P-1)$ path in
the subsequent time step. However, even 
in type-2 lossy equilibria, $P$-step oscillation
eventually resumes, e.g., at $t=2$ 
in~\cref{fig:equilibrium:loss:type-2}.
Type-1 equilibria typically appear for
a relatively high migration rate~$m$,
whereas type-2 equilibria tend to
appear for lower migration rates.

\subsection{Flow Equilibria}
\label{sec:lossy:flow}

While the lossy equilibria cannot be
characterized as simply
as the lossless equilibria in~\cref{sec:lossless:flow},
it is feasible to determine the flow-volume
bounds for the presented types of lossy equilibria.

Regarding the upper bound, the central question
is how high the upper boundary point~$\hat{f}^{\ast+}$
can become. Both for type-1 and type-2 lossy
equilibria, we note that~$\hat{f}^{\ast+}$
is reached after one round of $P$-step oscillation
starting from a flow volume below the capacity
limit. Hence, an upper bound on~$\hat{f}^{\ast+}$
can be represented as follows:
\begin{equation}
    \hat{f}^{\ast+} \leq \hat{f}_{\pi}^{(0)}(t_{\pi0}+P) \text{ where } \hat{f}_{\pi}^{(p)}(t_{\pi0}) = C/P.
     \label{eq:lossy:upper-bound}
\end{equation} The trajectory function
from~\cref{eq:flow:trajectory} (or from~\cref{eq:flow:trajectory:r1}
for~$r = 1$) is used to calculate the effects
of one round of $P$-step oscillation on 
flow volume~$C/P$, which is the highest
flow volume from which ordinary $P$-step oscillation
can proceed. Note that this trajectory
function is only usable if the agent dynamics
are in equilibrium according to~\cref{sec:lossless:agents}.
Type-1 lossy equilibria preserve~$P$-step
oscillation and thus also the corresponding agent 
equilibria. 
For type-2 lossy equilibria, however, $P$-step
oscillation is occasionally disturbed, which
can result in agent dynamics out of equilibrium.
However, multiple rounds
of~$P$-step oscillation precede 
the moment of reaching~$\hat{f}^{\ast+}$
and  the convergence to the agent
equilibrium is exponential. Hence, 
we observe that the agent dynamics are
close to the agent equilibrium and
the trajectory function can therefore be
used to obtain an approximate upper bound
for type-2 lossy equilibria.

Regarding the lower bound, we now investigate
how low the lower boundary point~$\hat{f}^{\ast-}$
can become. In type-1 lossy equilibria, $\hat{f}^{\ast-}$
is reached after $P-1$ time steps with
agent outflow on an overloaded rank-0 path.
Combined with the multiplicative decrease~$\beta$
in the first of these $P-1$ time steps,
we can thus formulate a lower bound 
on~$\hat{f}^{\ast-}$ for type-1 lossy equilibria:
\begin{equation}
   \hat{f}^{\ast-} > \beta\cdot(1-m)^{P-1}\cdot C_{\pi}
   \label{eq:lossy:lower-bound}
\end{equation}
given a rank-0 path that is only infinitesimally
overloaded and~$\alpha(\tau) > 0$, $\forall\tau \in \mathbb{N}_{>0}$.
For type-2 lossy equilibria, this lower bound
is too pessimistic, as the combination of multiplicative
decrease and agent out-migration directly transforms
the overloaded rank-0 path into the least utilized
path and there are no further consecutive
time steps with agent out-migration on this path.
Hence, $\beta\cdot(1-m)\cdot C_{\pi}$ suffices
as a lower bound for type-2 lossy equilibria.
For a validation of these lower bounds
by simulations, consult \cref{fig:lower-bound-validation}
in \cref{sec:appendix:additional-figures}.

\section{Axioms}
\label{sec:axioms}

In this section, we use an axiomatic approach
inspired by Zarchy et al.~\cite{zarchy2019axiomatizing}
to derive insights regarding the
effects of oscillatory path selection.
We adapt a number of their axioms, 
which were formulated for
a single-path context, to a
multi-path context in~\cref{sec:axioms:list}.
In~\cref{sec:axioms:equilibria},
we evaluate the equilibria 
from~\cref{sec:lossless,sec:lossy}
with respect to these axioms.

\subsection{List of Axioms}
\label{sec:axioms:list}

In our axiomatic approach to multi-path congestion control,
axioms correspond to desirable properties that
MPCC protocols should possess. 
However, as these properties refer to
general and vague concepts (e.g., efficiency or fairness), 
the conditions for possessing these properties
are usually not well-defined. Therefore,
the axioms here are formalized as metrics
for rating an MPCC protocol with respect to a certain
property, instead of binary indicators of whether
the protocol possesses the given property.
Concretely, we consider the following axioms
in this work:

\begin{axiom}
    \textbf{Efficiency.} An MPCC protocol is
    \textbf{$\boldsymbol{\epsilon}$-efficient} if under universal adoption
    of this protocol, the 
    bottleneck utilization
    of every path~$\pi$
    with capacity~$C/P$ is 
    never lower than a share $\epsilon$
    after some time~$t'$:
    \begin{equation}\exists t'.\quad \forall t \geq t',\ \pi\in\Pi.\quad \frac{P\cdot \hat{f}_{\pi}(t)}{C} \geq \epsilon
    \end{equation}
    Larger values of $\epsilon$ are better, and we consider an $\epsilon$-efficient protocol optimal if~$\epsilon \geq (C-s)/C$, where~$s$ is the buffer size.\label{ax:efficiency}\footnote{
    In terms of latency, $(C-s)/C$, i.e., empty buffers, would even
    be preferable to higher values of~$\epsilon$.
    This latency effect could be captured
    by an additional axiom, which we do 
    not introduce in this work.}
\end{axiom}

\begin{axiom}
    \textbf{Loss avoidance.}
    An MPCC protocol
    is~\textbf{$\boldsymbol{\lambda}$-loss-avoiding} if
    under universal adoption, 
    the loss rate on any path~$\pi$ with capacity~$C/P$ 
    never exceeds~$\lambda$ after some time~$t'$:
    \begin{equation}
        \exists t'.\quad \forall t > t',\ \pi\in\Pi.\quad \frac{\hat{f}_{\pi}(t)}{C/P} - 1 \leq \lambda
    \end{equation}
    Thus, smaller values of $\lambda$ are better, and a 0-loss-avoiding protocol is
    optimal.
    \label{ax:loss-avoidance}
\end{axiom}

\begin{axiom}
    \textbf{Convergence.}
    An MPCC protocol
    is~\textbf{$\boldsymbol{\gamma}$-convergent} if 
    under universal adoption, the flow volume~$\hat{f}_{\pi}(t)$ 
    on every path~$\pi$ lies
    consistently within a range $[\gamma\cdot \hat{f}_{\pi}^{+}, \hat{f}_{\pi}^{+}]$ 
    below a path-specific maximum level~$\hat{f}_{\pi}^{+}$
    after some time~$t'$:
\begin{equation}
    \exists t' > 0,\ f_{\pi}^{+}.\quad \forall t > t',\ \pi\in\Pi.\quad \gamma\cdot \hat{f}_{\pi}^{+} \leq \hat{f}_{\pi}(t) \leq \hat{f}_{\pi}^{+}
\end{equation}
    Thus, larger values of $\gamma$ are better, and a 1-convergent protocol is optimal.
    \label{ax:convergence}
\end{axiom}

\begin{axiom}
    \textbf{Fairness.} An MPCC protocol
    is~\textbf{$\boldsymbol{\eta}$-fair} 
    if under universal adoption,
    the  variance
    of congestion-window sizes 
    of all agents~$i \in A$ in the network
    never exceeds~$\eta$ after 
    some time~$t'$:\footnote{
    Zarchy et al.~\cite{zarchy2019axiomatizing}
    formalize fairness
    with the ratio of the smallest to the largest
    congestion-window size in the steady state.
    Given path selection, this ratio
    is always potentially 0, e.g., if an agent
    migrates in every time step.
    }
    \begin{equation}
        \exists t' > 0.\quad \forall t > t'.\quad \var_{i\in A}\big[\mathit{cwnd}_i(t)\big]\leq \eta
    \end{equation}
    \label{ax:fairness}
    Thus, smaller values of $\eta$ are better, and a 0-fair protocol is optimal.
\end{axiom}

For any axiom metric~$\mu$, we write
$\mu(\mathit{MPCC})$ for the most
desirable value of metric~$\mu$ that
the protocol~$\mathit{MPCC}$ can be 
rated with. 

% \begin{axiom}
%     \textbf{Responsiveness.} An MPCC protocol
%     is~\textbf{$\boldsymbol{m}$-responsive}
%     if the protocol has responsiveness parameter~$m$.
% \end{axiom}

\subsection{Axiomatic Characterization of Equilibria}
\label{sec:axioms:equilibria}

The axioms in~\cref{sec:axioms:list} refer to characteristics
which are eventually attained and then persistently preserved
by the flow dynamics. Hence, a natural way to axiomatically rate 
an MPCC protocol is to evaluate
the equilibria (i.e., stable states) of this protocol (cf.~\cref{sec:lossless,sec:lossy}).

\paragraph{Efficiency (\cref{ax:efficiency}).}
We distinguish lossless and lossy flow equilibria. If 
there is a lossless equilibrium 
($\hat{f}^{(0)} \leq C/P$), 
the minimal flow volume ever carried
by any path~$\pi$ is the equilibrium value for 
rank $P-1$, i.e.,~$\hat{f}^{(P-1)}$.\footnote{
To be precise, the asymptotic convergence
to~$\hat{f}^{(P-1)}$ permits that~$\hat{f}_{\pi}(t)$
for $\mathrm{rank}(\pi,t) = P-1$
is consistently below~$\hat{f}^{(P-1)}$.
However, since this shortfall is infinitesimal 
and flow volumes converge exponentially to their equilibrium value,
we treat the equilibrium as completely reached
instead of only asymptotically approached. 
} 
The network-wide efficiency level is 
therefore~$\epsilon = P\cdot\hat{f}^{(P-1)}/C$.
In contrast, for lossy equilibria, 
the efficiency level
is the lower bound on the lower boundary 
point~$\hat{f}^{\ast-}$ according to~\cref{sec:lossy:flow}.
Depending on the lossy-equilibrium type, 
this lower bound is
given by~$\beta\cdot(1-m)^{P-1}$
or~$\beta\cdot(1-m)$, respectively.
Since the lower bound is never higher
for type 1 than for type 2, we 
consider~$\beta\cdot(1-m)^{P-1}$
to be the minimum flow volume for lossy equilibria.
While this lower bound is too pessimistic for lossy
equilibria of type 2, these type-2 lossy equilibria
mostly appear for low values of~$m$, where the
difference between the two bounds is small.
\begin{equation}
    \epsilon\big(\mathit{MPCC}(\alpha,\beta, m, r)\big) \geq
    \begin{cases}
        P\cdot\hat{f}^{(P-1)}/C & \text{if } \hat{f}^{(0)} \leq C/P\\
        \beta \cdot (1-m)^{P-1} & \text{otherwise}
    \end{cases}
    \label{eq:ax:efficiency}
\end{equation}

\paragraph{Loss avoidance (\cref{ax:loss-avoidance}).}
If all paths are in lossless equilibrium 
($\hat{f}^{(0)} \leq C/P$), 
it is clear that the maximum loss
rate in the whole network is 0. 
If the network is in lossy equilibrium,
the maximum loss rate is determined by
the upper boundary point~$\hat{f}^{\ast+}$ 
(cf.~\cref{sec:lossy:flow}). As shown 
in~\cref{eq:lossy:upper-bound}, this boundary point
is maximal at~$\hat{f}_{\pi}^{(0)}(t_{\pi0}+P)$,
where~$\hat{f}_{\pi}^{(0)}$ is the rank-0 trajectory
function for an arbitrary path~$\pi$ and is anchored
at~$\hat{f}_{\pi}^{(0)}(t_{\pi0}) = C/P$.
For~$r \neq 1$ and~$r = 1$, this trajectory function
is given by~\cref{eq:flow:trajectory} 
and~\cref{eq:flow:trajectory:r1}, respectively.
In summary, the maximum loss rate is
\begin{equation}
    \lambda\big(\mathit{MPCC}(\alpha,\beta, m, r)\big) \leq 
    \begin{cases}
        0 & \text{if } \hat{f}^{(0)} \leq C/P\\
        \makecell{q(m,r,P)\cdot(1-m)^{P-1} - 1 +\\ \big(q(m,r,P)\cdot\sum_{p=0}^{P-2}\hat{\alpha}^{(p)}+\hat{\alpha}^{(P-1)}\big)\cdot\hat{a}^{(P-1)}/(C/P)} &  \makecell{\text{if } \hat{f}^{(0)} > C/P \\\land\ r \neq 1}\\
         \big((1-m)^{1-P}\cdot\sum_{p=0}^{P-2}\hat{\alpha}^{(p)}+\hat{\alpha}^{(P-1)}\big)\cdot \hat{a}^{(P-1)}/(C/P) & \text{otherwise},
    \end{cases}
    \label{eq:ax:loss-avoidance}
\end{equation} where we use the abbreviation $q(m,r, P):=(1+m\cdot r\cdot z(m,P))$.

\paragraph{Convergence (\cref{ax:convergence}).}
If the network is in a lossless equilibrium
($\hat{f}^{(0)} \leq C/P$), the convergence
behavior of the flow dynamics can be derived
from the boundaries~$\hat{f}^{(0)}$
and~$\hat{f}^{(P-1)}$ of the hypothetical flow
equilibrium. Given a lossy equilibrium,
we can build
on the range between the upper boundary
point~$\hat{f}^{\ast+}$ 
and the lower boundary
point~$\hat{f}^{\ast-}$, for which we have
derived an upper and a lower bound, respectively.
From these ranges, the derivation of the
convergence indicator~$\gamma$ and the maximum level~$\hat{f}_{\pi}^{+}$
is straightforward:
% \begin{subequations}
% \begin{equation}
%     \hat{f}_{\pi}^{+} = 
%     \begin{cases}
%         \hat{f}^{(0)} & \text{if }  \hat{f}^{(0)} \leq C/P\\
%         \big(\lambda(\mathit{MPCC}(\alpha,\beta,m,r)) + 1\big) \cdot C/P & \text{otherwise}
%     \end{cases}
% \end{equation}
\begin{equation}
    \gamma\big(\mathit{MPCC}(\alpha, \beta, m, r)\big) \geq
    \begin{cases}
        \hat{f}^{(P-1)}/\hat{f}^{(0)} & \text{if }  \hat{f}^{(0)} \leq C/P\\
        \beta\cdot(1-m)^{P-1}/\big(\lambda(\mathit{MPCC}(\alpha,\beta,m,r)) + 1\big) & \text{otherwise}
    \end{cases}
    \label{eq:ax:convergence}
\end{equation}
% \end{subequations}

\begin{figure}
\centering
\begin{minipage}{.48\textwidth}
  \centering
  \includegraphics[width=\linewidth,trim=0 0 10 0]{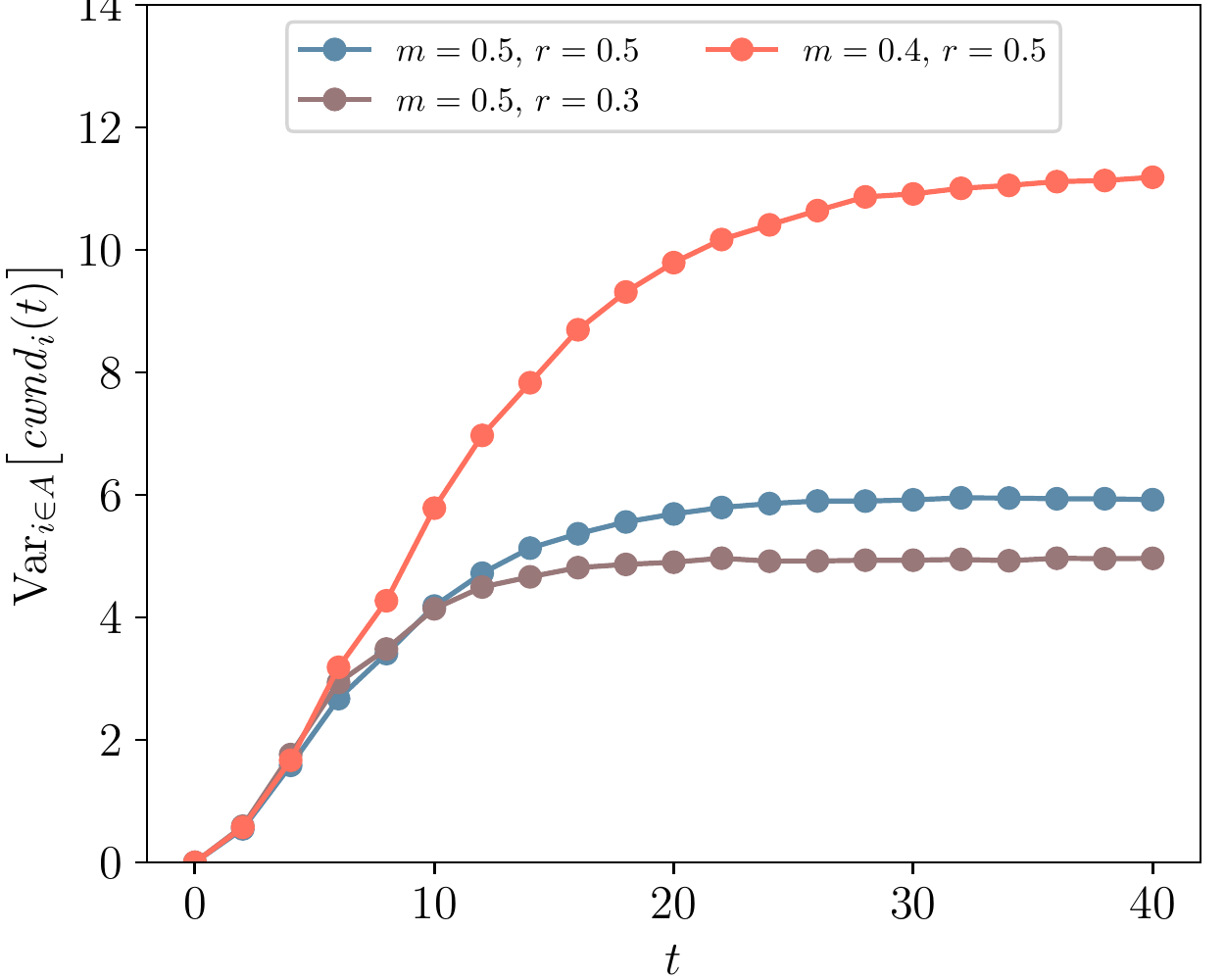}
  \caption{Computation of variance in congestion-window size
  according to the lossless Markov process in~\cref{eq:ax:fairness:lossless:markov}.}
  \label{fig:axioms:fairness:variance-computation}
\end{minipage}
\hfill
\begin{minipage}{.48\textwidth}
  \centering
  \begin{tikzpicture}[statesquare/.style={draw=black!50, shading=radial,inner color=white, thick, minimum width=10mm},
labelsquare/.style={fill=black, text=white, thick, minimum width=10mm},]

    % ------------------------------------------------
    % States
    
    % Increase
    \node[statesquare,align=center, outer color={rgb,255:red,255;green,200;blue,100}] (grow) at (0, 1.75) {%
        $\begin{aligned}
            \tau_i(t+1) &= \tau_i(t) + 1\\
            \mathit{cwnd}_i(t+1) &= \mathit{cwnd}_i(t) + \alpha(\tau_i(t))
        \end{aligned}$};
    \node[labelsquare] (growlabel) at (-2.7, 2.3) {\textbf{Increase}};
        
    % Migrate
    \node[statesquare,outer color={rgb,255:red,168;green,214;blue,245},align=center] (migrate) at (0,  0) {$\begin{aligned}
            \tau_i(t+1) &= 0\\
            \mathit{cwnd}_i(t+1) &= r\cdot\mathit{cwnd}_i(t)
        \end{aligned}$};
    \node[labelsquare] (migratelabel) at (-2.2, 0.55) {\textbf{Migrate}};
        
    % Loss
    \node[statesquare,align=center,outer color={rgb,255:red,214;green,245;blue,183}] (loss) at (0,  -1.75) {%
        $\begin{aligned}
            \tau_i(t+1) &= 0\\
            \mathit{cwnd}_i(t+1) &= \beta\cdot\mathit{cwnd}_i(t)
        \end{aligned}$};
    \node[labelsquare] (decreaselabel) at (-2.3, -1.2) {\textbf{Decrease}};
        
    % ------------------------------------------------
    % Transitions
    
    % Grow -> Migrate
    \draw[-{Triangle}] ($(grow.south) + (0.1, 0)$) -- ($(migrate.north) + (0.1, 0)$);
    \node at ($(grow.south) + (0.6, -0.3)$) {$p_{I\rightarrow M}$}; %{$m$};
    
    % Migrate -> Grow
    \draw[-{Triangle}] ($(migrate.north) - (0.1, 0)$) -- ($(grow.south) - (0.1, 0)$);
    \node at ($(grow.south) + (-0.6, -0.3)$) {$\overline{p_{\ell}}\cdot\overline{m}$};
    
    % Migrate -> Loss
    \draw[-{Triangle}] ($(migrate.south) + (0.1, 0)$) -- ($(loss.north) + (0.1, 0)$);
    \node at ($(migrate.south) + (0.8, -0.25)$) {$p_{\ell} \cdot \overline{m}$};
    
    % Loss -> Migrate
    \draw[-{Triangle}] ($(loss.north) - (0.1, 0)$) -- ($(migrate.south) - (0.1, 0)$);
    \node at ($(migrate.south) + (-0.5, -0.3)$) {$m$};
    
    %  Grow -> Grow
    \draw[-{Triangle}] ($(grow.west) + (0, 0.2)$) to [out=180,in=140]  ($(grow.west) - (0, 0.2)$);
    \node[align=left] at ($(grow.west) + (-0.55, 0)$) {$p_{I\rightarrow I}$}; % $\overline{p_{\ell}}\cdot\overline{m}$};
    
    % Migrate -> Migrate
    \draw[-{Triangle}] ($(migrate.west) + (0, 0.2)$) to [out=180,in=140]  ($(migrate.west) - (0, 0.2)$);
    \node at ($(migrate.west) + (-0.4, 0)$) {$m$};
    
    % Grow -> Loss
    \draw[-{Triangle}] ($(grow.south) + (2.2, 0)$) to [out=270,in=80] ($(loss.east) + (0, 0.1)$);
    \node at ($(grow.south) + (1.7, -0.3)$) {$p_{I\rightarrow D}$}; %{$p_{\ell} \cdot \overline{m}$};
    
    % Loss -> Grow
    \draw[-{Triangle}] ($(loss.east) - (0, 0.1)$) to [out=70,in=270] ($(grow.south) + (2.4, 0)$);
    \node at ($(loss.east) + (0.25, -0.35)$) {$\overline{m}$};

\end{tikzpicture}
  \caption{Markov process for congestion-window
  size given lossy equilibria (Notation: $\overline{p} = 1-p$).
  Moreover, $p_{I\rightarrow I} = \text{if }  
  \tau_i(t) \bmod P \neq (P-1) \text{ then } \overline{p_{\ell}}\cdot\overline{m} \text{ else } \overline{p_{\ell}}$, $p_{I\rightarrow M} = \text{if } \tau_i(t) \bmod P \neq (P-1)
  \text{ then } m \text{ else } 0$ and
  $p_{I\rightarrow D} = \text{if } \tau_i(t) \bmod P \neq (P-1)
  \text{ then } p_{\ell}\cdot\overline{m} \text{ else } p_{\ell}$.
%   Note that
%   \textit{Increase}, \textit{Migrate}
%   and \textit{Decrease} do not designate states,
%   but transitions.
  }
  \label{fig:axioms:fairness:lossy:markov}
\end{minipage}
\end{figure}

\paragraph{Fairness (\cref{ax:fairness}).}
We consider the variance of congestion-window sizes
in the equilibrium as a metric for the fairness
of an MPCC algorithm:
\begin{align}
    \var_{i\in A}\big[\mathit{cwnd}_i(t)\big] &=  \expectation_{i\in A}\big[\mathit{cwnd}_i(t)^2\big] - \expectation_{i\in A}\big[\mathit{cwnd}_i(t)\big]^2
\end{align}

As the congestion-window evolution of a single
agent is a probabilistic process where any state
transition only depends on the current state,
we approximate~$\mathit{cwnd}_i(t)$ for the case 
of lossless equilibria by means of
the following Markov process with two state variables:
\begin{equation}
\begin{aligned}
        &\mathbf{if} \ \tau_i(t) \neq P-1\text{:} &\tau_i(t+1),\ \mathit{cwnd}_i(t+1) =
        &\begin{cases}
            0,\ r \cdot \mathit{cwnd}_i(t) & \text{prob. $m$}\\
            \tau_i(t) + 1,\ \mathit{cwnd}_i(t) + \alpha(\tau_i(t))& \text{prob. $1 - m$}
        \end{cases}\\
        &\mathbf{else\text{:}} &\tau_i(t+1),\ \mathit{cwnd}_i(t+1) =\ &\tau_i(t)+1,\ \mathit{cwnd}_i(t) + \alpha(\tau_i(t))
    \label{eq:ax:fairness:lossless:markov}
\end{aligned}\end{equation} where the
initial state is given by~$\tau_i(0) = \mathit{cwnd}_i(0) = 0$.

Computationally tractable computation of
the congestion-window size variance
can be done by averaging many simulation samples 
of the Markov process 
from~\cref{eq:ax:fairness:lossless:markov},
which has only linear complexity in~$t$
and yields the expectation of the 
congestion-window size by the central limit theorem.
\cref{fig:axioms:fairness:variance-computation}
illustrates that the variance
of~$\mathit{cwnd}_i(t)$ has a limit for~$t\rightarrow\infty$.

Regarding lossy equilibria, the Markov process
from~\cref{eq:ax:fairness:lossless:markov} must be
adapted as shown in~\cref{fig:axioms:fairness:lossy:markov}.
In particular, we assume that every path encounters loss
with probability~$p_{\ell}$ in any time step, except
if the path has experienced loss in the previous time step
(as there are no consecutive loss events on the same path
in the lossy equilibria in~\cref{sec:lossy}). If the agent 
is using a lossy path, but does not leave the path, 
the congestion-window size is multiplicatively decreased
as shown in transition \textit{Decrease} 
in~\cref{fig:axioms:fairness:lossy:markov}.
Like for lossless equilibria, a simulation-based approach
enables to efficiently compute the variance in congestion-window size
(cf.~\cref{fig:axioms:fairness:variance:lossy}
in the appendix). This figure suggests that the
variance limit for lossy equilibria is decreasing 
in loss probability~$p_{\ell}$.
Moreover, since the lossy Markov process 
in~\cref{fig:axioms:fairness:lossy:markov}
is equivalent to the lossless Markov process
in~\cref{eq:ax:fairness:lossless:markov} 
for~$p_{\ell} = 0$, the variance of the
lossless Markov process represents
an upper bound on the variance of the lossy
Markov process. Therefore, we henceforth 
exclusively rely on the lossless Markov process.
\section{Axiom-Based Insights}
\label{sec:insights}

In this section, we derive fundamental insights
into the nature of end-host path selection
on the basis of the axioms presented
in the previous section. First, we
investigate in~\cref{sec:insights:effects} 
how the performance characteristics
of a network change if end-host path selection is
introduced. Second, we show in~\cref{sec:insights:trade-off}
that there are fundamental trade-offs
when applying end-host path selection.

\subsection{Performance Effects of Introducing End-Host Path Selection}
\label{sec:insights:effects}

\subsubsection{Evaluation Method}

In order to analyze how end-host path selection
affects the performance characteristics of a
network, we use a comparative approach: 
First, we characterize the performance
of a network without end-host path selection
based on the axioms from~\cref{sec:axioms:list} 
(henceforth: Scenario~(I)). 
Afterwards,
we compare the axiomatic ratings of
the network without path selection
to the axiomatic characterization
of the MPCC equilibria 
(cf.~\cref{sec:axioms:equilibria})
that arise in the same network
given end-host path selection
(henceforth: Scenario~(II)).

We base the comparison on a
network with~$N$ agents
and a total bottleneck
capacity~$C$ distributed over~$P$
paths with
equal bottleneck capacity~$C/P$.
All agents adopt
the same CC protocol~$\mathit{CC}_i(\alpha,\beta)$
(cf.~\cref{eq:cc-protocol})
in Scenario~(I),
whereas they employ a multi-path 
version~$\mathit{MPCC}_i(\alpha,\beta,m,r)$
of this CC protocol in Scenario~(II). 
Moreover, while the
agent distribution on paths is
dynamically determined in Scenario~(II),
the agent distribution
in Scenario~(I) is static:
To identify the worst-case effects
of end-host path selection, let
this static agent distribution
be optimal, 
i.e., $a_{\pi} = N/P$.

In the following, we rate the 
CC~protocol~$CC_i(\alpha,\beta)$ for
Scenario~(I) with respect to the
axioms and perform a comparison with
Scenario (II). Moreover,
we both quantify and interpret
the changes in the axiom metrics
that are due to the introduction
of end-host path selection.
These changes are also visualized
in~\cref{fig:effects:efficiency-loss,fig:effects:convergence-fairness}: For any~$m$ and every
equilibrium class (lossless or lossy),
the possible range of the metric change 
is shown for two different
additive-increase functions and an example
network. We distinguish a constant 
additive-increase function~$\alpha_1(\tau) = 1$
and an additive-increase function~$\alpha_{\mathrm{S}}$ in the style
of TCP Slow Start:~$\alpha_{\mathrm{S}}(\tau) = 2^{\tau}$
if~$\tau < 5$ and $\alpha_{\mathrm{S}}(\tau) = 1$ otherwise. 
The range associated with each value of~$m$
is $[\min_{r \in R(m)} \Delta(m,r),\ \max_{r \in R(m)} \Delta(m,r)]$, where~$\Delta$ is the difference
metric as a function of~$m$ and~$r$ and
~$R(m)$ contains all
values of~$r$ that produce a valid
equilibrium of the given class (lossless or lossy) 
in the example network given~$m$.

\subsubsection{Efficiency (\cref{ax:efficiency}).}
Given that the employed protocol~$CC_i(\alpha,\beta)$
eventually exhausts the capacity of any path,
the efficiency level is given by the 
lowest possible flow volume that results
from loss. This lower bound is determined
by the multiplicative decrease~$\beta$
applied to a flow volume that is 
infinitesimally above the capacity limit:
\begin{equation}
    \forall \pi\in\Pi.\ \epsilon\big(\mathit{CC}_i(\alpha,\beta)\big) = \frac{\beta\cdot C/P}{C/P} = \beta
\end{equation}

\begin{figure}
\centering
\begin{subfigure}{.48\textwidth}
  \centering
  \includegraphics[width=\linewidth,trim=0 10 10 10]{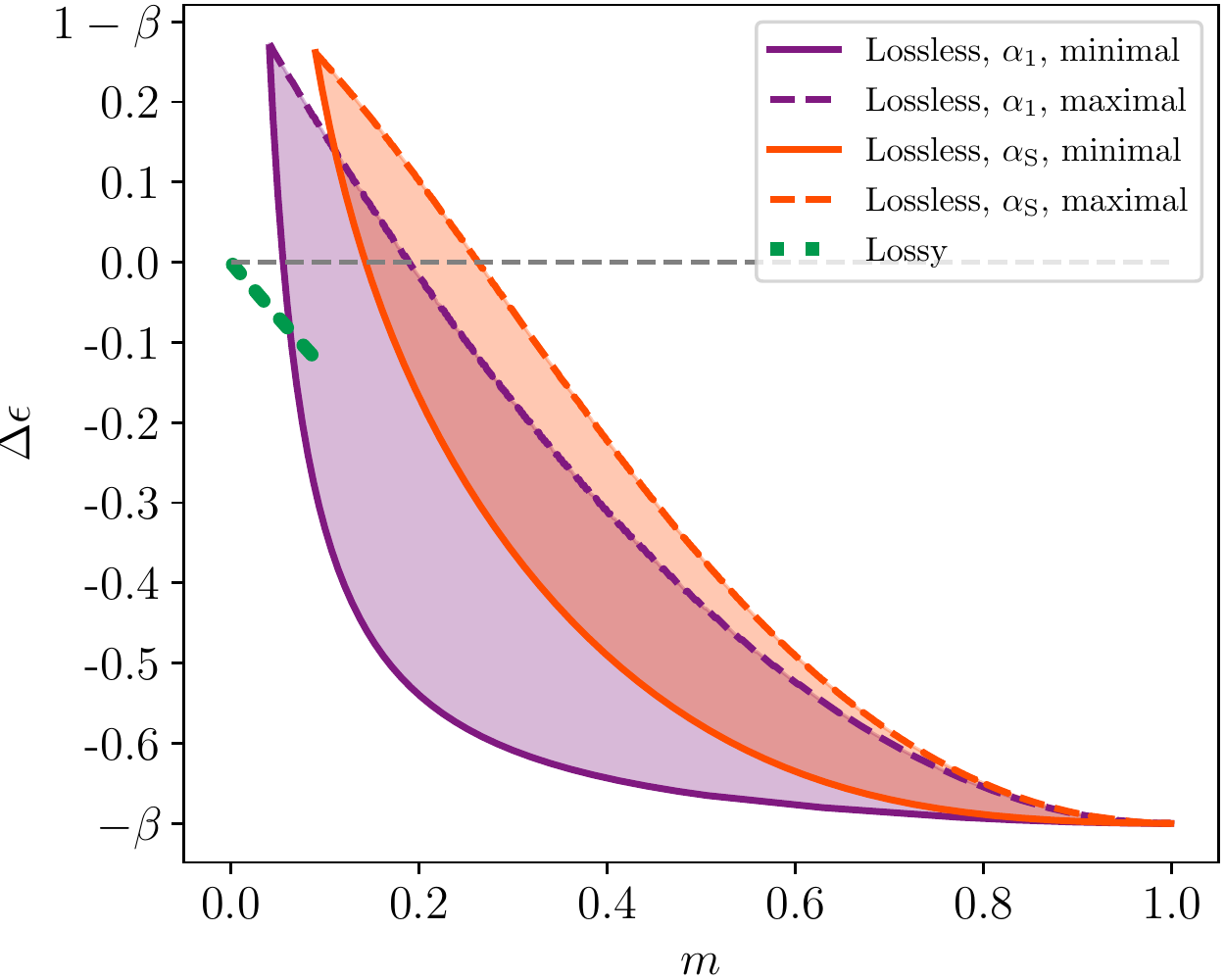}
  \caption{Change in efficiency level~$\epsilon$.}
  \label{fig:effects:efficiency}
\end{subfigure}%
\hfill
\begin{subfigure}{.48\textwidth}
  \centering
  \includegraphics[width=\linewidth,trim=10 10 0 10]{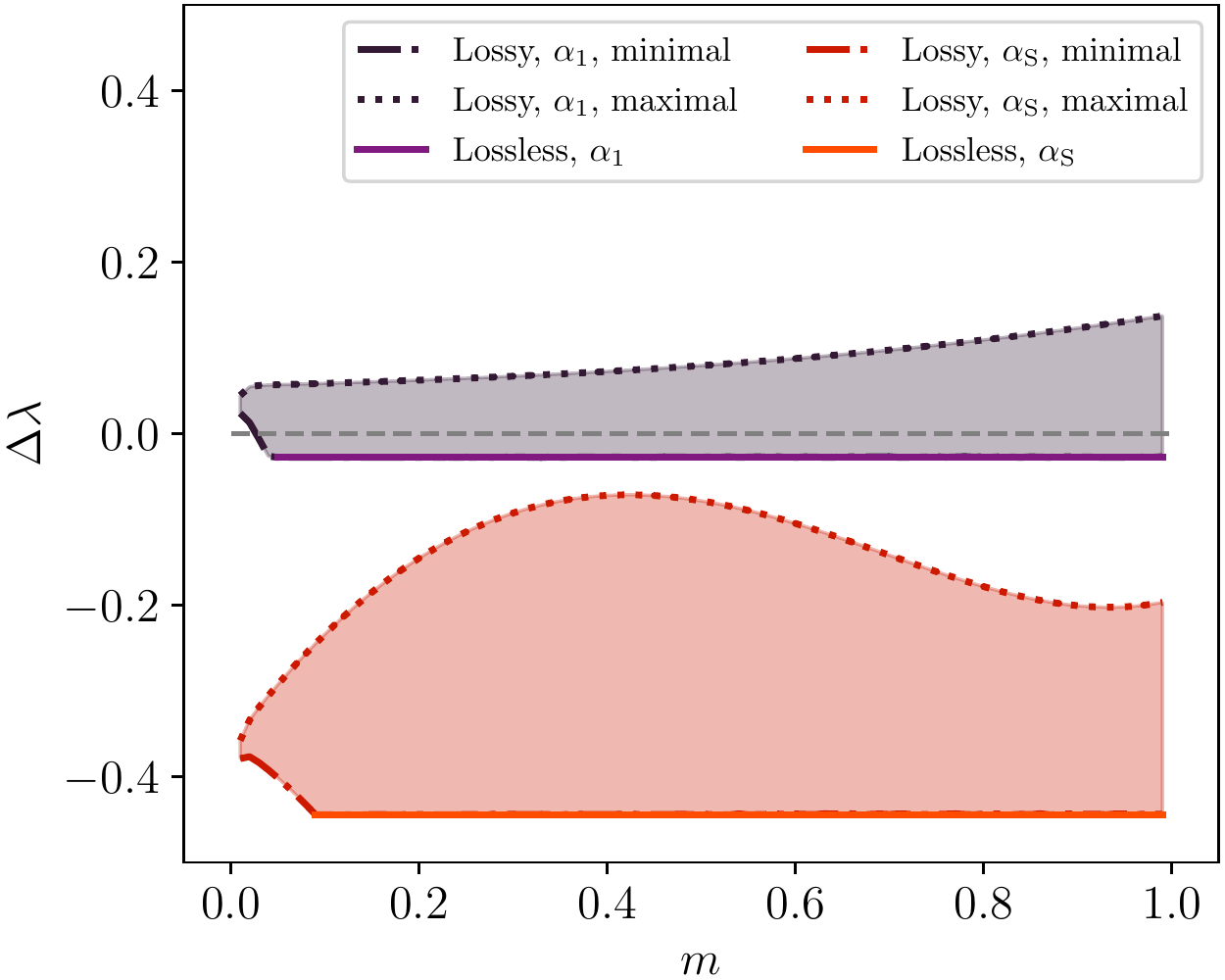}
  \caption{Change in loss level~$\lambda$.}
  \label{fig:effects:loss}
\end{subfigure}
\caption{Effects of end-host
path selection for an example network 
with $P=3$, $\beta=0.7$, $N=1000$, and
$C/P = 12000$.}
\label{fig:effects:efficiency-loss}
\end{figure}

We now compare this efficiency level
to the MPCC efficiency levels
from~\cref{eq:ax:efficiency}
and analyze the efficiency
change~$\Delta\epsilon = \epsilon\big(\mathit{MPCC}(\alpha,\beta,m,r)\big)-\epsilon\big(\mathit{CC}(\alpha,\beta)\big)$
that is due to the introduction
of end-host path selection.
For a visualization of this efficiency
change, consider~\cref{fig:effects:efficiency}.

If the efficiency level of the MPCC
dynamics is determined by a lossless
equilibrium, then~$\Delta\epsilon$
is given by~$P\cdot \hat{f}^{(P-1)}/C - \beta$.
As~$\hat{f}^{(P-1)}$ is a decreasing
function of the migration rate~$m$
and an increasing function of
the reset softness~$r$, end-host
path selection is more likely to
negatively affect~$\epsilon$
for high migration rates and hard resets
on path switch: 
\begin{insight}
    \textbf{Efficiency Effects of
    Path Migration and Resets in Lossless Equilibria.}
    The more likely agents
 are to migrate away from a path at
 any single point in time,
the further the bottleneck-link
utilization can drop, and
if agents start out with a small
congestion window every time they switch
to a new path, utilization (and therefore efficiency) are relatively low.
\label{ins:efficiency:migration-reset}
\end{insight}
Nonetheless, it is possible
that the introduction of end-host path selection 
leads to a higher level of efficiency.
The computations for the example network,
visualized in~\cref{fig:effects:efficiency},
show that for low values of~$m$ and high
values of~$r$, introducing end-host
path selection can increase efficiency.

In contrast, if the MPCC efficiency
level is determined by a lossy equilibrium,
then~$\Delta\epsilon$ is given by
$\beta\cdot(1-m)^{P-1} - \beta$,
which is bound to be negative.
This fact allows the interpretation
that end-host path selection strictly
lowers the efficiency in case of loss,
as emigration from a path reinforces
the utilization plunge created by the CC loss
reaction, i.e., the multiplicative decrease~$\beta$.
As~\cref{fig:effects:efficiency} shows,
such less efficient lossy equilibria are bound to
exist for low values of~$m$, for which
there is no value of~$r$ such
that a a lossless equilibrium
can arise. This insight points
to a subtle relationship between migration
rates and efficiency: 
\begin{insight}
    \textbf{Inefficient Equilibria due to Low Migration.}
    While lowering the migration
rate can increase the efficiency of end-host
path selection, very low migration rates
necessarily lead to inefficient (lossy) equilibria,
which make
end-host path selection detrimental
to efficiency compared to a scenario
without path selection.
\label{ins:efficiency:low-migration}
\end{insight} 

\subsubsection{Loss avoidance (\cref{ax:loss-avoidance}).}
In Scenario~I, the worst-case loss rate occurs if
flow~$f_{\pi}$ on path~$\pi$ is exactly 
at the capacity limit~$C_{\pi}$, and
there is an additional increase
by all agents on the path:
\begin{equation}
    \forall \pi\in\Pi.\ \lambda\big(\mathit{CC}_i(\alpha,\beta)\big) =
    \frac{\alpha^{\max}\cdot a_{\pi}}{C_{\pi}}
    = \frac{\alpha^{\max}\cdot N}{C},
\end{equation} where~$\alpha^{\max} = \max_{\tau \in \mathbb{N}_{\geq 0}} \alpha(\tau)$ to represent
the maximum possible loss.

In case of lossless equilibria
of the MPCC dynamics, it is clear
that~$\Delta\lambda$ (defined analogously 
to~$\Delta\epsilon$) is negative, i.e.,
the loss rate can be reduced (to 0).
This improvement in~$\Delta\lambda$ is
shown in~\cref{fig:effects:loss}
for all values of~$m$ for which there
is a value of~$r$ such that a 
lossless equilibrium arises.

If a lossy equilibrium is present,
the effects of end-host path selection
are more ambivalent. In that case,
the maximum loss rate in the path-aware network
is proportional to~$\hat{f}^{(0)}$: the larger
the hypothetical limit value~$\hat{f}^{(0)}$ of 
the trajectory function,
the stronger the increase of the trajectory function
at level~$C_{\pi}$ and thus the higher the loss rate.
As~$\hat{f}^{(0)}$
is proportional to~$r$ and effectively
infinite for~$r = 1$, the highest loss rate
for every value of~$m$ is achieved for~$r = 1$, 
which yields the following intuitive insight: 
\begin{insight}
    \textbf{Loss Effects of Soft Resets.}
    If agents only perform
soft resets of the congestion-window size 
when switching paths, this can
result in high loss on the newly
selected patt.
\label{ins:loss:reset}
\end{insight}
In contrast, if~$m$ and~$r$ are
such that the equilibrium is only
marginally lossy, i.e., $\hat{f}^{(0)}$
is only infinitesimally larger than $C_{\pi}$,
then the maximum loss rate in a lossy equilibrium
is arbitrarily close to 0 (similar to a lossless
equilibrium). However, a value of~$r$ that
achieves~$\hat{f}^{(0)} \approx C_{\pi}$ may not
exist given a (low) value of~$m$, in which
case the reduction of the loss rate to~0
is not possible. Therefore, we arrive
at a counter-intuitive insight that
mirrors~\cref{ins:efficiency:low-migration}:
\begin{insight}
    \textbf{Loss Effects of Low Migration.}
    Loss is not minimized by minimizing
the migration rate~$m$, as low migration
rates may prohibit the emergence
of completely lossless equilibria.
\label{ins:loss:low-migration}
\end{insight}

Furthermore,~\cref{fig:effects:loss}
allows another non-obvious insight:
\begin{insight}
    \textbf{Loss Effects of Path Selection
    with Variable Additive-Increase Functions.}
    The benefits of end-host path selection
in terms of loss are particularly large
if additive-increase functions with high
inherent variability (such as~$\alpha_{\mathrm{S}}$
in~\cref{fig:effects:efficiency-loss})
are used by the agents.
\label{ins:loss:continuity-time}
\end{insight}
In that case, end-host path selection
may reduce loss because
it leads to de-synchronization of the continuity
time~$\tau$ between agents: If all agents
tend to have the same continuity time~$\tau$,
there is a chance that many agents have continuity
time~$\tau_{\max}$ with~$\alpha(\tau_{\max}) = \alpha^{\max}$
at the same time, resulting in high loss. 
In contrast, agent migration due to path selection
causes more heterogeneity in~$\tau$ and
therefore leads to an averaging of~$\alpha(\tau)$,
which reduces the aggregate additive increase
and therefore the maximum possible loss.
While this observation may first
seem like an unfair comparison of a
maximum to an average, the averaging
of additive increases is exactly the
fundamental feature of path selection
that reduces the possible maximum
of aggregate additive increase 
compared to a scenario without path selection.

\subsubsection{Convergence (\cref{ax:convergence}).}
The convergence level~$\gamma$ is determined
by the minimum and
the maximum possible flow volume, as derived
above:
\begin{equation}
    \gamma\big(\mathit{CC}_i(\alpha,\beta)\big)
    = \frac{\beta\cdot C}{C + \alpha^{\max} \cdot N}
\end{equation}

% Depending on whether a lossless or a lossy equilibrium
% determines the convergence level of the MPCC dynamics,
% the change in the stability level due to end-host
% path selection is:
% \begin{equation}
%     \Delta\gamma = \frac{\hat{f}^{(P-1)}}{\hat{f}^{(0)}} - \frac{\beta\cdot C}{C+\alpha^{\max}\cdot N} \quad \text{or} \quad
%     \Delta\gamma = \frac{\beta\cdot(1-m)^{(P-1)}\cdot C_{\min}}{(1+m\cdot r\cdot z)\cdot C_{\min} + \hat{\alpha}^{(P-1)}\cdot\hat{a}^{(P-1)}} - \frac{\beta\cdot C}{C+\alpha^{\max}\cdot N}
% \end{equation}

\begin{figure}
\centering
\begin{subfigure}{.48\textwidth}
  \centering
  \includegraphics[width=\linewidth,trim=0 0 10 10]{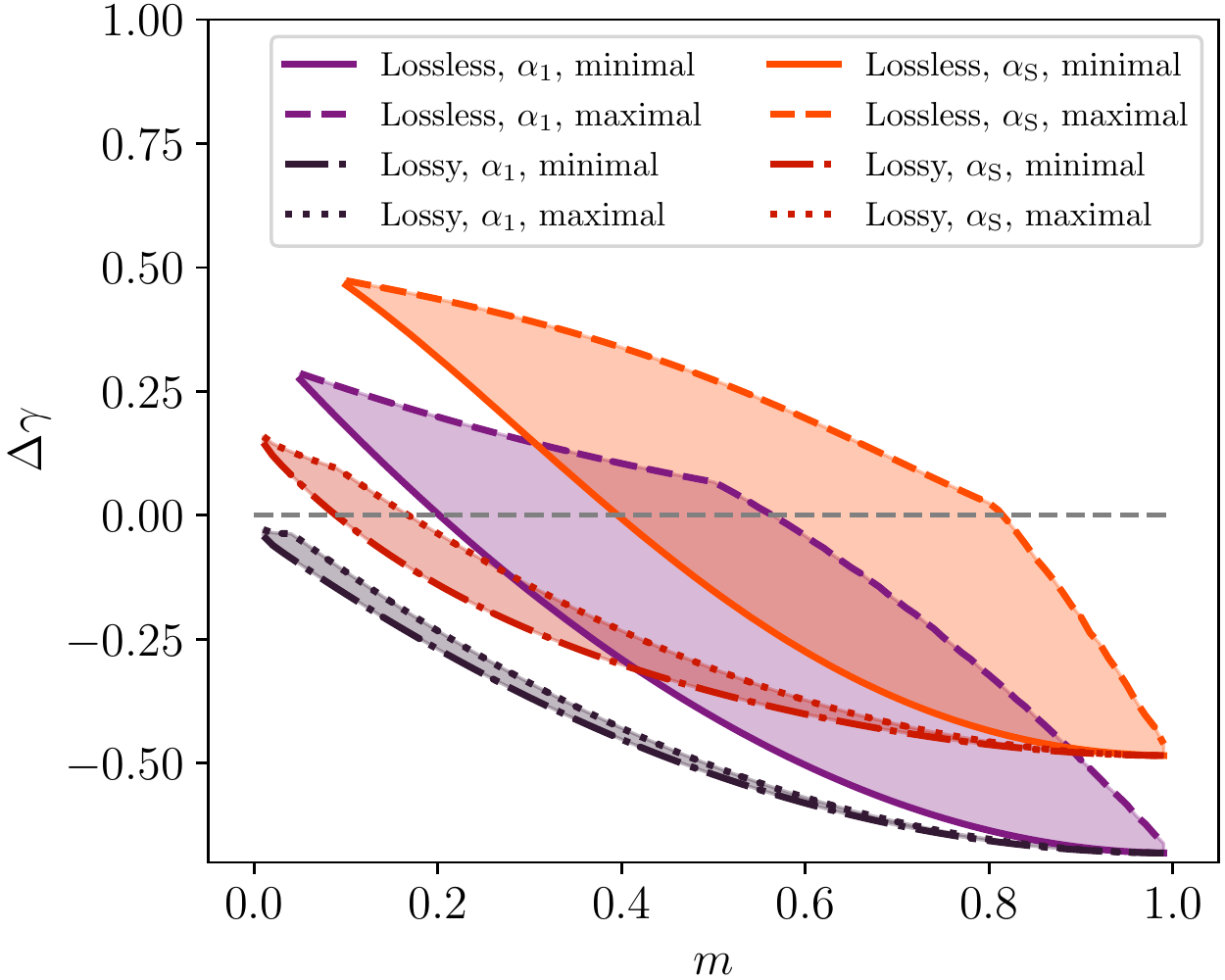}
  \caption{Change in convergence level~$\gamma$.}
  \label{fig:effects:convergence}
\end{subfigure}%
\hfill
\begin{subfigure}{.48\textwidth}
  \centering
  \includegraphics[width=\linewidth,trim=10 0 0 10]{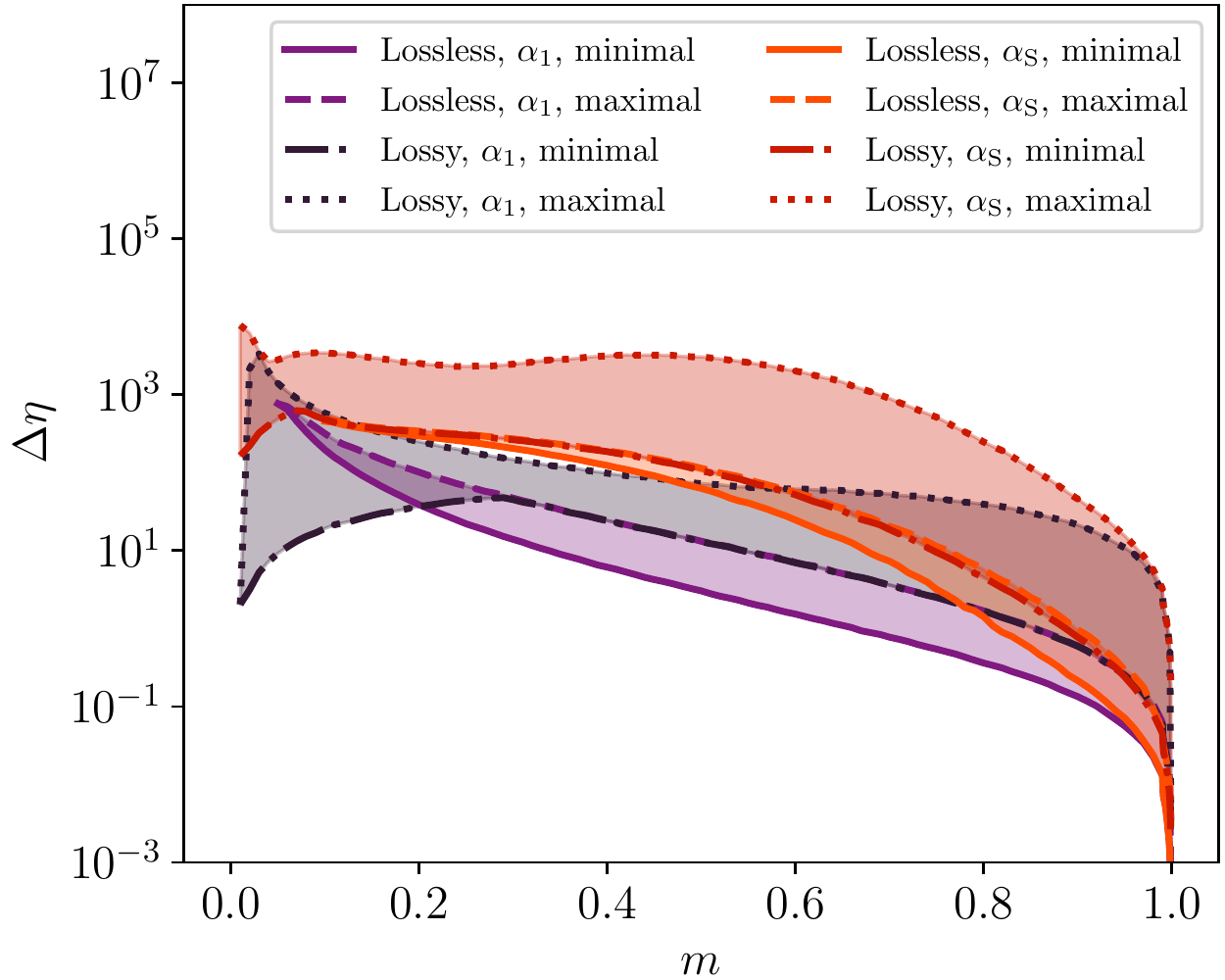}
  \caption{Change in fairness level~$\eta$.}
  \label{fig:effects:fairness}
\end{subfigure}
\caption{Illustration of effects of end-host
path selection on the basis of the same
example network as in~\cref{fig:effects:efficiency-loss}.}
\label{fig:effects:convergence-fairness}
\end{figure}

In the case of lossless equilibria,
end-host path selection can increase
stability if
\begin{equation}
    \frac{\hat{f}^{(P-1)}}{\hat{f}^{(0)}} =
    \frac{\sum_{p=0}^{P-2}\hat{\alpha}^{(p)} +
    \hat{\alpha}^{(P-1)}\cdot(1-m)^{P-1}}
    {\big(1+m\cdot r\cdot z(m,P)\big)\cdot\sum_{p=0}^{P-2}\hat{\alpha}^{(p)} +
    \hat{\alpha}^{(P-1)}} > \gamma\big(\mathit{CC}_i(\alpha,\beta)\big),
    \label{eq:insights:delta-gamma}
\end{equation} which is unsurprisingly
true for a low migration rate~$m$
and hard resets~$r \approx 0$. However,
analogously to efficiency and loss, 
convergence surprisingly suffers from 
very low migration rates~$m$, as
this causes lossy 
equilibria,
which are inferior to lossless equilibria
in terms of convergence (cf.~\cref{fig:effects:convergence}).

The convergence in these lossy equilibria
benefits from low migration rates
and hard resets, without the exception
for very low migration rates that exists
for lossless equilibria. While such lossy equilibria
might be inferior to lossless equilibria
in terms of convergence, lossy equilibria 
of end-host path selection might still
be preferable to no end-host path selection
at all, as~\cref{fig:effects:convergence} suggests
for the lossy equilibria for~$\alpha_{\mathrm{S}}$.
Similar to~\cref{ins:loss:continuity-time}, 
the reason for this improvement is
the de-synchronization of the continuity time
brought about by agent migration, which
reduces the variance of the aggregate
additive increase and thus the flow-volume fluctuations.
Contrary to the widespread belief that
end-host path selection necessarily hurts stability
(in the sense of the convergence axiom),
our analysis thus shows that network stability
can in fact benefit from end-host path selection.

\subsubsection{Fairness (\cref{ax:fairness}).}
Given simultaneous sending start 
and no path selection, perfect 
synchronization implies that all
agents always have exactly the 
same congestion-window size, 
i.e., ~$\eta = 0$.
Moreover, Zarchy et al. show
that even if some agents start sending
after others, CC protocols generally tend to
come close to perfect fairness~\cite{zarchy2019axiomatizing}.
To find the worst-case effects
of end-host path selection, we
thus assume perfect fairness in
the scenario without path selection:
\begin{equation}
    \eta\big(\mathit{CC}_i(\alpha,\beta)\big)
    = 0
\end{equation}

Hence, the fairness change~$\Delta\eta$ due to 
end-host path selection is equal to the 
fairness level~$\eta$ of the MPCC dynamics,
which has been computed as a function of the
migration rate~$m$ in~\cref{fig:effects:fairness}
for two different additive-increase functions.
In~\cref{fig:effects:fairness}, 
the lowest values for~$\eta$, i.e.,
the highest fairness, is achieved
for very high migration rates~$m \approx 1$,
which leads to the following insight:
\begin{insight}
    \textbf{Fairness Effects of Path Migration.}
    In a system with end-host path-selection,
    a very high migration rate~$m$ leads to
    optimal fairness.
\end{insight}
This phenomenon can be intuitively explained
as follows: If the
migration rate is high, any agent is likely
to reset its congestion-window size in any
time step, which results in a
compact distribution of the congestion-window size.
Under a low
migration rate, some agents may reach
a high congestion-window size due to
uninterrupted growth,
while a few agents per time step perform a reset, which leads to a high
variance of the congestion-window
size distribution.

The effects of the reset softness~$r$
on~$\eta$ are more nuanced.
As~\cref{fig:effects:fairness} shows,
the fairness metric~$\eta$ is generally
higher for lossy equilibria, which
appear for high reset softness,
than for lossless equilibria, i.e.,
lossless equilibria are fairer.
However, as mentioned in~\cref{sec:axioms:equilibria},
the fairness metric for lossy equilibria
is computed for infinitesimal 
loss probability~$p_{\ell}$; for any 
higher~$p_{\ell}$,~$\eta$ is lower,
which complicates a comparison to lossless
equilibria. Also, for a low migration rate~$m$,
lossy equilibria with a high reset softness 
are associated with lower~$\eta$ 
than lossless equilibria. The reason
behind this phenomenon is that
soft resets reduce the difference
in congestion-window size between
the agents that have not migrated
in a long time (and therefore have
a large congestion-window) and the
agents that have just recently migrated
and reset their congestion-window size.

Finally, while end-host path selection
seems to reduce fairness as captured
by~$\eta$, we note that~$\eta$ only
represents the fairness at any
single point in time. However,
under low migration rates, there
may still be very high \emph{inter-temporal}
fairness. If the migration probability
is low, any agent has a high probability
to uninterruptedly grow its congestion window
for a long time. If the 
congestion-window sizes of any agent 
were averaged over a certain
time span, the distribution of such
average congestion-window sizes would have
low variance. We leave this more complex
fairness analysis as an interesting
task for future work.

\subsection{Fundamental Trade-Offs}
\label{sec:insights:trade-off}

In~\cref{sec:axioms:equilibria,sec:insights:effects}, the
dependency of the MPCC dynamics on
the migration rate (or responsiveness)~$m$ and the 
reset softness~$r$ has been qualified and 
quantified. These characterizations allow
to observe the following trade-off 
in the design of systems with end-host path selection:
\begin{insight}
    \textbf{Trade-Off Regarding
    Migration Rate.}
    Efficiency~$\epsilon$
    and convergence~$\gamma$
    are more favorable under \textbf{low} migration rates,
    whereas fairness~$\eta$ and responsiveness~$m$
    are more favorable under \textbf{high} migration rates,
    implying a fundamental trade-off between these axioms.
    \label{ins:trade-off:migration}
\end{insight}

However, we note that this trade-off is only valid
\emph{within} equilibrium classes, e.g., for comparing 
lossless equilibria among each other, but not \emph{across}
equilibrium classes: Lowering the migration rate below a
certain (low) level restricts the set of possible
equilibria to lossy equilibria, which are worse
in terms of efficiency and convergence than %the
lossless equilibria 
% that were possible for higher
% migration rates
(cf.~\cref{ins:efficiency:low-migration}).

Regarding loss avoidance, the effect of
migration rates depends on the remaining network parameters.
If resets are hard ($r\approx 0$), higher migration
rates are associated with lower loss rate
(as higher migration rates make lossless equilibria
more likely, which are optimal in terms of loss).
In contrast, if resets are soft ($r \approx 1$),
lossless equilibria are impossible and
the effects of the migration rate on the
loss rate are unclear in general, because
the migration rate non-monotonically affects 
the aggregate additive increase 
(cf. the curve for maximal~$\Delta\lambda$
given~$\alpha_{\mathrm{S}}$
and lossy equilibria in~\cref{fig:effects:loss}).
However, this unpredictable effect vanishes
for constant additive-increase functions
(such as~$\alpha_1$ from~\cref{sec:insights:effects}).
For constant-increase
functions, a higher migration rate
leads to a \emph{higher} loss rate given soft
resets. This finding underlines
the relevance of congestion-window
adaptation on path switch: Depending
on the reset softness, higher migration
rates may increase or reduce loss.

Despite this subtle relationship
of migration rates and the axiomatic metrics,
we can identify parameters~$m$ and~$r$ 
that are optimal with respect 
to all the metrics efficiency,
loss, and convergence simultaneously.
These parameters are given by
the lowest~$m$ such that a lossless
equilibrium is still possible given
a complete reset~$r = 0$.
These parameters yield
a lossless equilibrium with high efficiency
and convergence (cf.~\cref{ins:trade-off:migration}).
\begin{insight}
    \textbf{No Trade-Off between Efficiency, Loss Avoidance
    and Convergence.}
    Since there exist
    protocol parameters that are optimal
    with respect to efficiency,
    loss avoidance, and convergence
    simultaneously, there exists no
    fundamental trade-off between these
    metrics.
\end{insight} Unfortunately, determining these
optimal parameters requires
knowledge about specific and variable
properties of the target network, i.e.,
the number of agents~$|A|$ and the 
path-bottleneck 
capacities~$C_{\pi}$ in the network,
making it unattainable in most practical
settings.

Finally, when determining the reset softness~$r$,
a further trade-off arises:
\begin{insight}
    \textbf{Trade-Off Regarding
    Reset Softness.}
    There is a fundamental trade-off between convergence~$\gamma$
    and loss~$\lambda$, both of which
    are more favorable under \textbf{low} reset softness (hard resets),
    and efficiency~$\epsilon$, 
    which is more favorable under \textbf{high} reset softness (soft resets).
\end{insight}

\section{Related Work}
\label{sec:related-work}

Traditionally, the effects of end-host path selection
have been theoretically studied in the literature
on \emph{selfish routing}. In this line of research,
the classic Wardrop 
model~\cite{wardrop1952road}
is used to characterize stable traffic distributions
(equilibria) that result from uncoordinated 
path selection by self-interested agents.
These equilibria have been thoroughly investigated
with respect to their 
existence~\cite{rosenthal1973class, roughgarden2007routing}, 
their efficiency~(typically
termed \emph{Price of Anarchy}~\cite{koutsoupias1999worst, roughgarden2002bad, roughgarden2003price, qiu2003selfish}), 
and their convergence properties
\cite{fischer2009adaptive,sandholm2001potential,fischer2004evolution}.
However, the Wardrop model cannot represent congestion-control dynamics
appropriately, which we consider important
for characterizing the impact of end-host
path selection on network performance.

In research about multi-path congestion control,
there has been widespread use of fluid models
which can better represent congestion-control
dynamics~~\cite{peng2013multipath, key2007path, kelly2005stability, han2006multi}.
However, also these models focus on representing
equilibria in terms of approximate traffic distributions
on networks and do not capture stability-relevant small-scale dynamics
such as congestion-window fluctuations.
% Exactly these fluctuations are relevant
% for stability, which is a
% concern in networks with end-host path selection.
More applied approaches rely on
reasoning from network examples and
experimental validation and have
been used in the design of MPTCP algorithms
such as LIA~\cite{wischik2011design,raiciu2011coupled} 
and OLIA~\cite{khalili2013mptcp}. These
approaches
% capture such fluctuations better, but
are rather suited for the design of
concrete protocols than for the elicitation
of fundamental properties of end-host path selection.
Moreover, MPTCP research typically only investigates
the effects of path selection by scrutinizing
friendliness concerns between single-path
and multi-path TCP users in the same network, 
not by looking at the impact
that the introduction of end-host path selection
has on aggregate performance based on various metrics.

In contrast, the axiomatic approach used in this paper
allows to qualify and quantify the performance
impact of path selection on a fundamental
level while taking congestion-control dynamics into account. 
% One further strength of the axiomatic approach consists of
% its ability to identify and describe the
% relationships between multiple relevant metrics.
Thanks to this power, the axiomatic perspective
has been applied to various topics beyond 
game theory: In computer science,
for example, research on congestion 
control~\cite{zarchy2019axiomatizing}, 
routing protocols~\cite{lev2016axiomatic},
and recommendation systems~\cite{andersen2008trust}
have benefited from axiom-based approaches.

The effects of end-host path selection have also been
characterized by Wang et al.~\cite{wang2011cost},
whose `cost of not splitting in routing' captures
the difference in network utility between a scenario
where end-hosts select a single path and a scenario
where multiple paths can be selected. However,
this work differs from ours in investigating
static rate allocations instead of dynamic
rate evolution, 
in evaluating a single metric (utility) instead
of multiple axioms, and in contrasting
different modes of end-host path selection
instead of contrasting path selection with path 
pre-determination.
\section{Conclusion}
\label{sec:conclusion}

Motivated by a stability concern about
end-host path selection, 
we qualify and quantify the performance
impact of such path-selection-induced instability 
in this work. More precisely, we analyze a general
network in which end-hosts employ greedy load-adaptive
path selection and characterize the
resulting traffic pattern with respect to
five metrics of interest (``axioms''): efficiency,
loss avoidance, convergence, fairness and
responsiveness.
Through this analysis, we show how the performance
impact of end-host path selection depends
on the path-migration behavior, the underlying
congestion-control protocol, and the structure
of the network. Among the dependencies that we present
and explain, there are both intuitive,
well-known dependencies (e.g., high migration
rates decrease efficiency) and non-intuitive,
more complex dependencies (e.g., 
very low migration rates increase loss).
Moreover, we show that there are 
fundamental limitations
such that no multi-path congestion-control
protocol can optimize all metrics simultaneously.

We understand our work as a first step,
which allows many avenues for follow-up
research. For example, it would
be interesting to extend the model
for additional congestion-control
behaviors (e.g., latency-based protocols
or model-based protocols such as 
BBR~\cite{cardwell2016bbr}), additional
path-switching behaviors (e.g.,
based on a path-switching probability
proportional to the load difference
between paths) 
and more general networks.
However, while our insights admittedly stem
from a simplified model, we believe
that the illustrated dependencies
and the axiomatic reasoning in general
can inform the discussion about
the merits and perils of
end-host path selection.

\section*{Acknowledgements}

We gratefully acknowledge support from ETH Zurich, 
from SNSF for project ESCALATE (200021L\_-182005),
and from the Austrian Science Fund (FWF) 
for project I 4800-N (ADVISE), 2020-2023.
Moreover, we thank Joel Wanner for his helpful
feedback, the anonymous reviewers for their 
careful reviews, and Edmundo de Souza e Silva 
for his shepherding.

\bibliographystyle{ACM-Reference-Format}
\bibliography{references}

\newpage
\appendix

\section{Analysis of the Continuity-Time Distribution}
\label{sec:equilibria:distribution}

The agent dynamics involved in $P$-step oscillation (\cref{def:p-step}) 
allow to estimate how long the agents on a path have already 
been using that path without a packet loss, 
i.e., allow to characterize the distribution of the \emph{continuity time}
introduced above. For the following analysis,
we introduce the notation~$\theta(\pi, t)$,
which shall denote the time since  the most recent loss event 
on path~$\pi$ at time~$t$.

We now derive a probability 
distribution~$\mathbb{P}\big[\tau_{i}(t) = \tau\big]$, 
giving the probability
that agent~$i$ on path~$\pi_i(t)$ has continuity 
time~$\tau \in \mathbb{N}$ at time~$t$.
This distribution will later be used to determine the expected
congestion-window increase~$\hat{\alpha}_\pi(t)$
in~\cref{eq:flow:expected}.
We consider an arbitrary 
agent~$i \in A$ at an arbitrary 
time~$t$, residing on path~$\pi_i(t)$.
Clearly, agent~$i$ must have continuity time~$\tau_{i}(t) = 0$
right after a loss event, i.e., 
whenever~$\theta(\pi_i(t), t) = 0$, 
irrespective of the rank of~$\pi_i(t)$:
\begin{equation}
    \forall t \text{ s.t. } \theta\big(\pi_i(t), t\big) = 0,\ p \in [P].\quad
    \mathbb{P}\left[\tau_i(t) = 0\ \middle|\ \mathrm{rank}\big(\pi_i(t), t\big) = p\right] = 1
\end{equation}

However, in the subsequent
time steps, where~$\theta\big(\pi_i(t), t\big) > 0$, 
the continuity-time distribution of agent~$i$ 
on path~$\pi_i(t)$ depends on the rank of that path.
If~$\mathrm{rank}(\pi_i(t),t) = 0$
or, equivalently,~$\mathrm{rank}(\pi_i(t-1), t-1\big) = P-1$,
all the~$\hat{a}^{(P-1)}$ agents
that were on path~$\pi$ in the last time step~$t-1$ 
have remained on the path and 
have increased their continuity time by 1,
but their \emph{relative share} is reduced by
on-migration from other paths:
 \begin{equation}\begin{aligned}
   &\forall t \text{ s.t. } \theta\big(\pi_i(t), t\big) > 0,\ \tau > 0.\\
   &\mathbb{P}\left[\tau_{i}(t) = \tau\ \middle|\ \mathrm{rank}\big(\pi_i(t), t\big) = 0 \right]\\
    &= \mathbb{P}\left[\tau_{i}\big(t-1\big) = \tau-1\ \middle|\ \mathrm{rank}\big(\pi_i(t-1), t-1\big) = P-1 \right]\cdot \hat{a}^{(P-1)}/\hat{a}^{(0)}\\
    &= \mathbb{P}\left[\tau_{i}\big(t-1\big) = \tau-1\ \middle|\  \mathrm{rank}\big(\pi_i(t-1), t-1\big) = P-1 \right]\cdot(1-m)^{P-1}
\end{aligned}\end{equation}
All the~$m\cdot (N - \hat{a}^{(P-1)})$ agents
that migrated from the other paths have continuity time~$0$:
\begin{equation}
    \forall t \text{ s.t. } \theta\big(\pi_i(t), t\big) > 0.\quad \mathbb{P}\left[\tau_{i}(t) = 0\ \middle|\  \mathrm{rank}\big(\pi_i(t), t\big) = 0 \right]
    = \frac{m\cdot (N-\hat{a}^{(P-1)})}{\hat{a}^{(0)}} = 1-(1-m)^{P-1}.
\end{equation}
If path~$\pi_i(t)$ has $\mathrm{rank}(\pi_i(t), t) \neq 0$, 
the continuity-time distribution has
been shifted up by 1 in the last time step, 
but is otherwise unaffected:
\begin{equation}\begin{aligned}
    &\forall t \text{ s.t. } \theta\big(\pi_i(t), t\big) > 0,\ p \in [P]\setminus\{0\},\ \tau > 0.\\ &\mathbb{P}\big[\tau_{i}(t) = \tau\ |\ \mathrm{rank}\big(\pi_i(t), t\big) = p \big]
    = \mathbb{P}\big[\tau_{i}(t-1) = \tau - 1\ |\ \mathrm{rank}\big(\pi_i(t-1), t-1\big) = p-1 \big]
\end{aligned}\end{equation}

\begin{figure}
\centering
\begin{minipage}{.48\textwidth}
  \centering
  \includegraphics[width=\linewidth,trim=0 0 10 0]{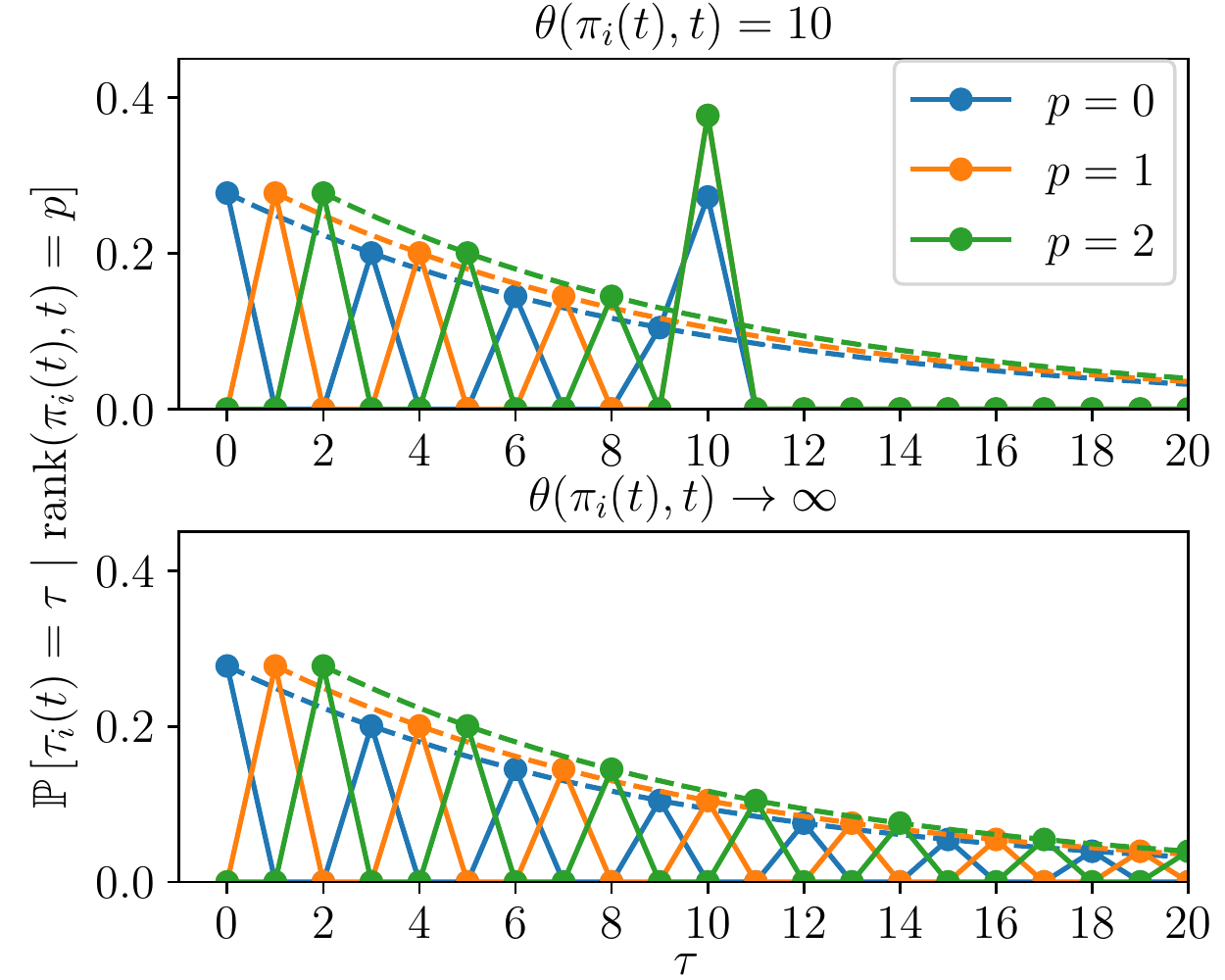}
  \caption{Continuity-time distribution
  for $P = 3$, $m = 0.15$, and different~$\theta$.
  The dashed lines represent the function~$(1-(1-m)^{P-1})\cdot(1-m)^{(\tau-p)/P\cdot(P-1)}$.}
  \label{fig:continuity-time}
\end{minipage}
\hfill
\begin{minipage}{.48\textwidth}
  \centering
  \includegraphics[width=\linewidth,trim=10 0 0 0]{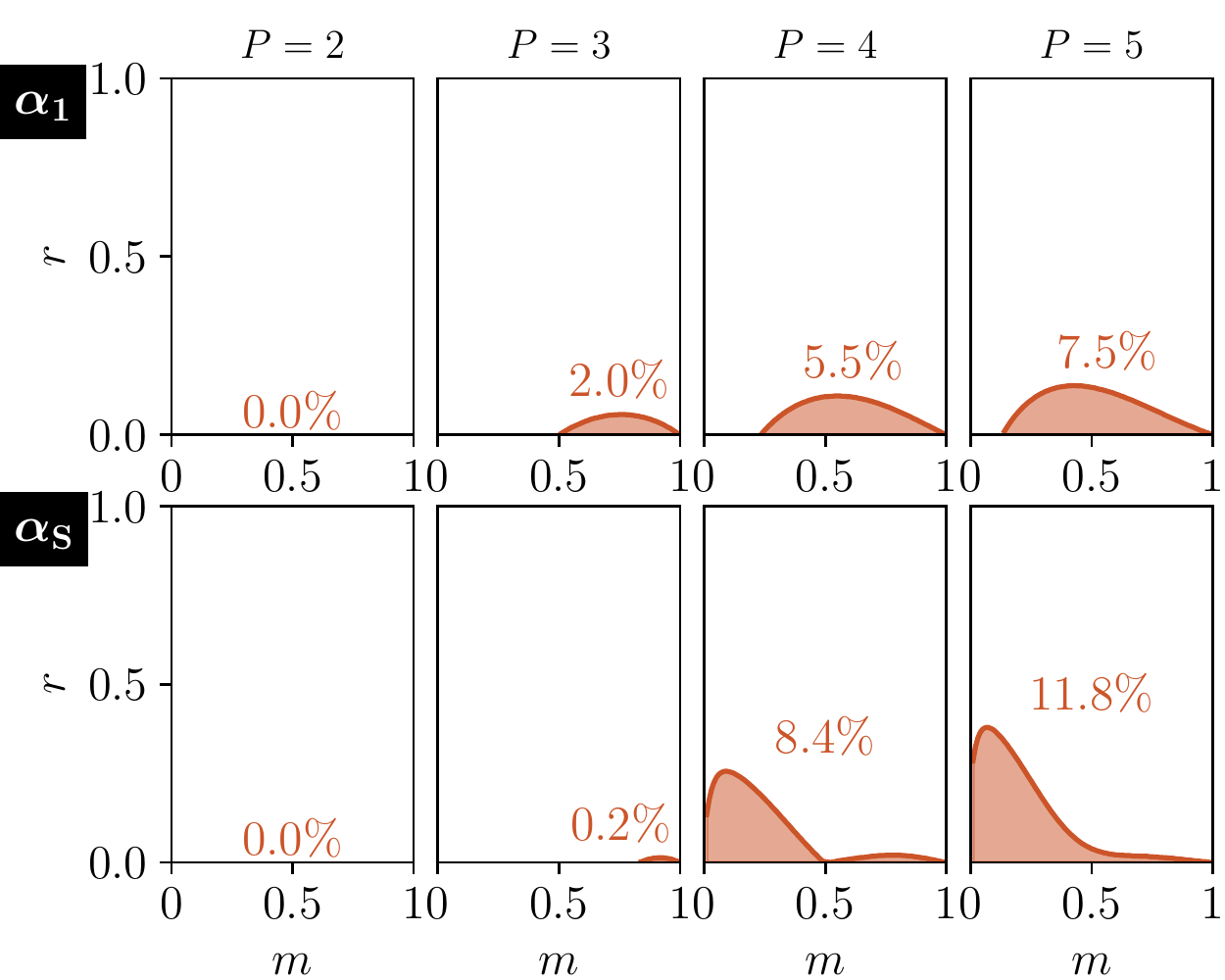}
  \caption{Visualization of parameter sub-space that is inconsistent with~$P$-step oscillation
  for different additive-increase 
  functions~$\alpha_1(\tau) = 1$ and
  $\alpha_{\mathrm{S}}(\tau) = (2^{\tau}
  \text{ if } \tau < 5 \text{ else } 1)$.}
  \label{fig:p-step-oscillation-consistency}
\end{minipage}
\end{figure}

These recursive characterizations of the probability distribution
are equivalent to the following explicit definition of the
continuity-time distribution,
which is visualized
in~\cref{fig:continuity-time}:

\begin{insight}
    At time~$t$, the probability that an
    agent~$i$ on a path~$\pi_i(t)$ with rank $p$ and time since last loss~$\theta\big(\pi_i(t), t\big)$ has continuity time~$\tau$ is 
    \begin{equation}
    \begin{split}
        \mathbb{P}\big(\tau; t, p\big) &:= \mathbb{P}\big[\tau_{i}(t) = \tau\ |\ \mathrm{rank}\big(\pi_i(t), t\big) = p\big]\\
        &\hphantom{:}= \begin{cases}(1-m)^{\left\lceil\frac{\tau-p}{P}\right\rceil(P-1)} & \text{if $\tau = \theta\big(\pi_i(t), t\big)$}\\
        (1-(1-m)^{P-1})\cdot(1-m)^{\left\lfloor\frac{\tau}{P}\right\rfloor(P-1)} & \text{if } \tau < \theta\big(\pi_i(t), t\big) \land \tau \bmod P = p\\
        0 & \text{otherwise}\end{cases}
    \end{split}
    \end{equation}
    \label{ins:continuity-time}
\end{insight}

On a path~$\pi$ with rank~$p$ at time~$t$, the expected additive 
increase~$\hat{\alpha}_\pi(t)$
at time~$t$ is therefore:
\begin{align}
    \hat{\alpha}_\pi(t)
    % &= \mathbb{E}\big[\alpha(\tau_{\pi i}(t_{\pi}^{(p)}))\ |\ \tau_{\pi\ell}\big]
    % \\
    &= (1-m)^{\left\lceil\frac{\theta-p}{P}\right\rceil(P-1)}\cdot\alpha(\theta) +
    \sum_{k = 0}^{\lceil(\theta-p)/P\rceil - 1} (1-(1-m)^{P-1})\cdot(1-m)^{k(P-1)} \cdot \alpha(Pk + p)
    \label{eq:exp-alpha:theta}
\end{align} where~$\theta = \theta(\pi, t)$.

For increasing time since the last loss 
($\theta \rightarrow \infty$),
the expected average
additive increase 
on a path with rank~$p$ converges to the following quantity,
which can be easily computed for any additive-increase
function~$\alpha$:
\begin{align}\label{eq:alpha-p}
    \hat{\alpha}^{(p)} = \sum_{k = 0}^{\infty} (1-(1-m)^{P-1})\cdot(1-m)^{k(P-1)} \cdot \alpha(Pk+p)
\end{align}

\section{Approximation Accuracy}
\label{sec:appendix:approximation-accuracy}

\begin{figure}
    \centering
    \includegraphics[width=\linewidth]{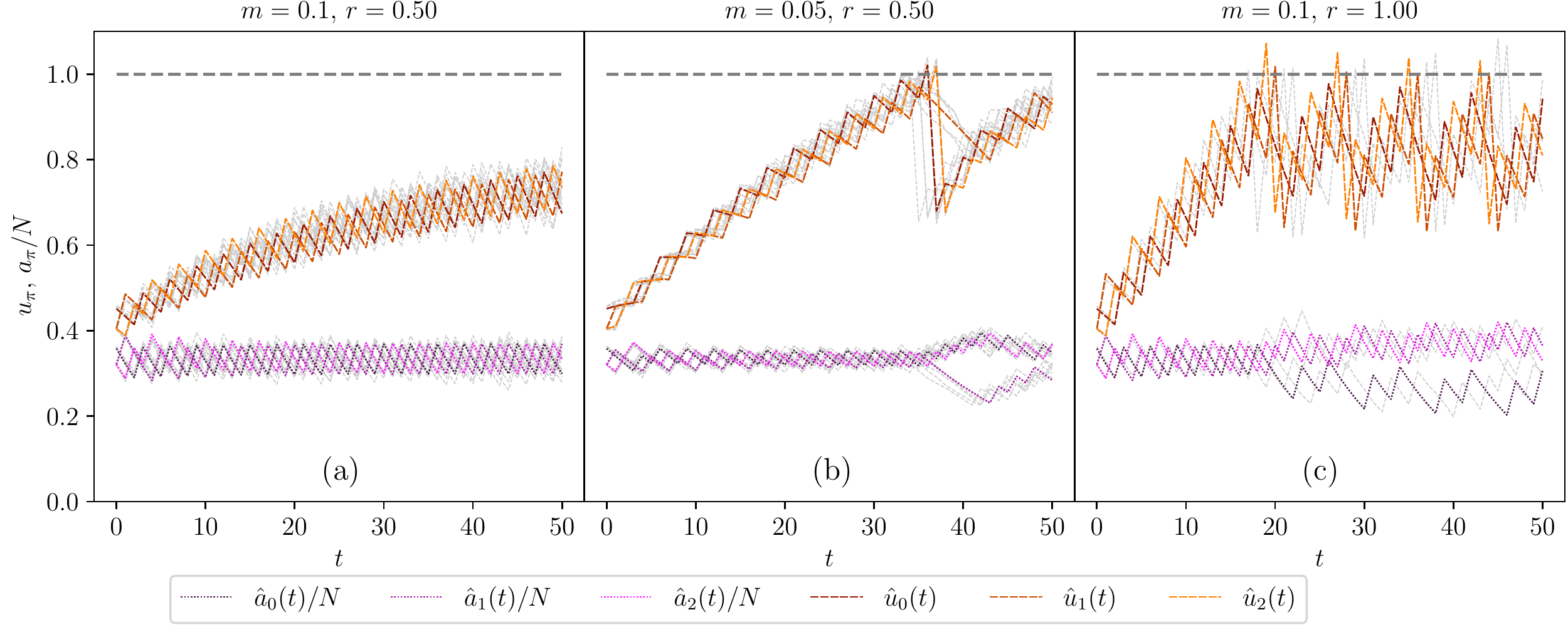}
    \caption{Simulated MPCC dynamics $\{(a_{\pi}(t),\ f_{\pi}(t))\}_{\pi\in\Pi}$ (gray dashed)
    and expected MPCC dynamics $\{(\hat{a}_{\pi}(t),\ \hat{f}_{\pi}(t))\}_{\pi\in\Pi}$ (in color) for
    $P =3$, $N = 1000$, $\alpha(\tau) = 1$, $\beta = 0.7$, and $C_{\pi} = 12\,000$ for every~$\pi \in \Pi$.}
    \label{fig:approximation-accuracy}
\end{figure}

The expected MPCC dynamics in~\cref{eq:expected} are an approximation
of the actual probabilistic MPCC dynamics in~\cref{eq:raw},
which are unsuitable for analytic investigation.
In order to demonstrate the accuracy of this approximation,
we present a comparison between the actual dynamics and
the expected dynamics for a selection of parameters
in~\cref{fig:approximation-accuracy}.
In each sub-figure, the actual
MPCC dynamics from \cref{eq:agents:raw,eq:flow:raw} 
are simulated and shown with light gray
lines, and the expected dynamics are computed
and drawn with colored lines
(agent dynamics in dotted lines,
flow dynamics in dashed lines).

In \cref{fig:approximation-accuracy}(a) and
\cref{fig:approximation-accuracy}(b) (i.e., $r\neq1$),
the expected dynamics are compared with results
from 5 simulation runs of the actual dynamics.
The expected dynamics appropriately capture
the structure of both the agent dynamics
and the flow dynamics, in particular
the curvature, the convergence behavior
and the reaction to loss (e.g., at $t = 35$
in~\cref{fig:approximation-accuracy}(b)).
As the actual dynamics are realizations
of a random variable, their values
deviate from the expectation; however, the variance
is modest.

In~\cref{fig:approximation-accuracy}(c) (i.e, $r=1$),
the actual flow dynamics look more different from
the expected flow dynamics than for $r\neq1$.
This difference is due to loss events
at different points in time, which can even result
in case of low variance,
but make the dynamics look quite different.
However, the pattern of recurring loss
is well captured by the expected dynamics.
In order to make this similarity visible,
only one simulation run of the actual dynamics
is shown.

\begin{figure}
    \centering
    \begin{subfigure}{\linewidth}
        \centering
        \includegraphics[width=0.85\linewidth]{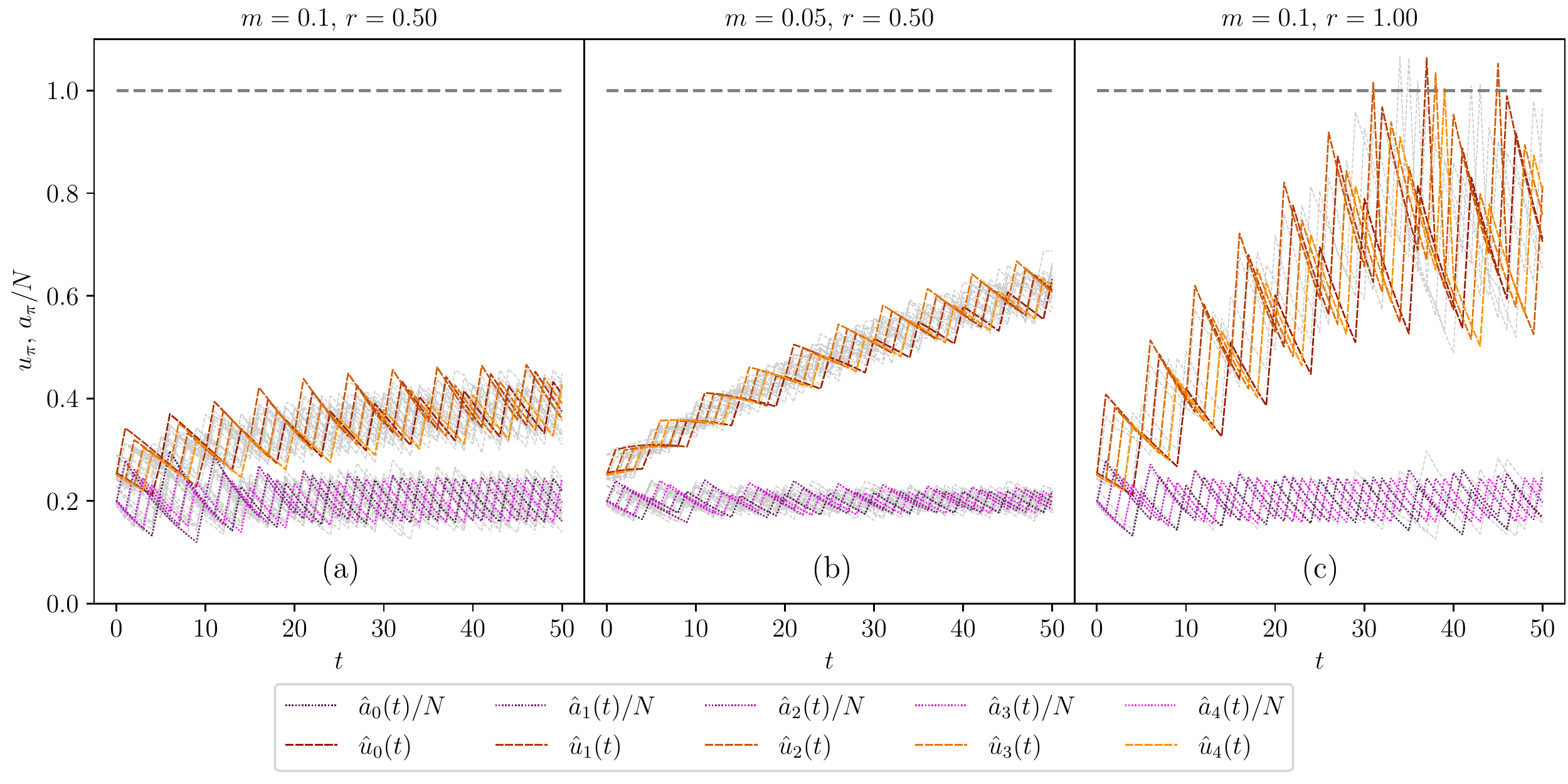}
        \caption{$P=5$, $\alpha_1(\tau) = 1$.}
        \label{fig:approximation-accuracy:P5:CONSTANT}
    \end{subfigure}
    \begin{subfigure}{\linewidth}
        \centering
        \includegraphics[width=0.85\linewidth]{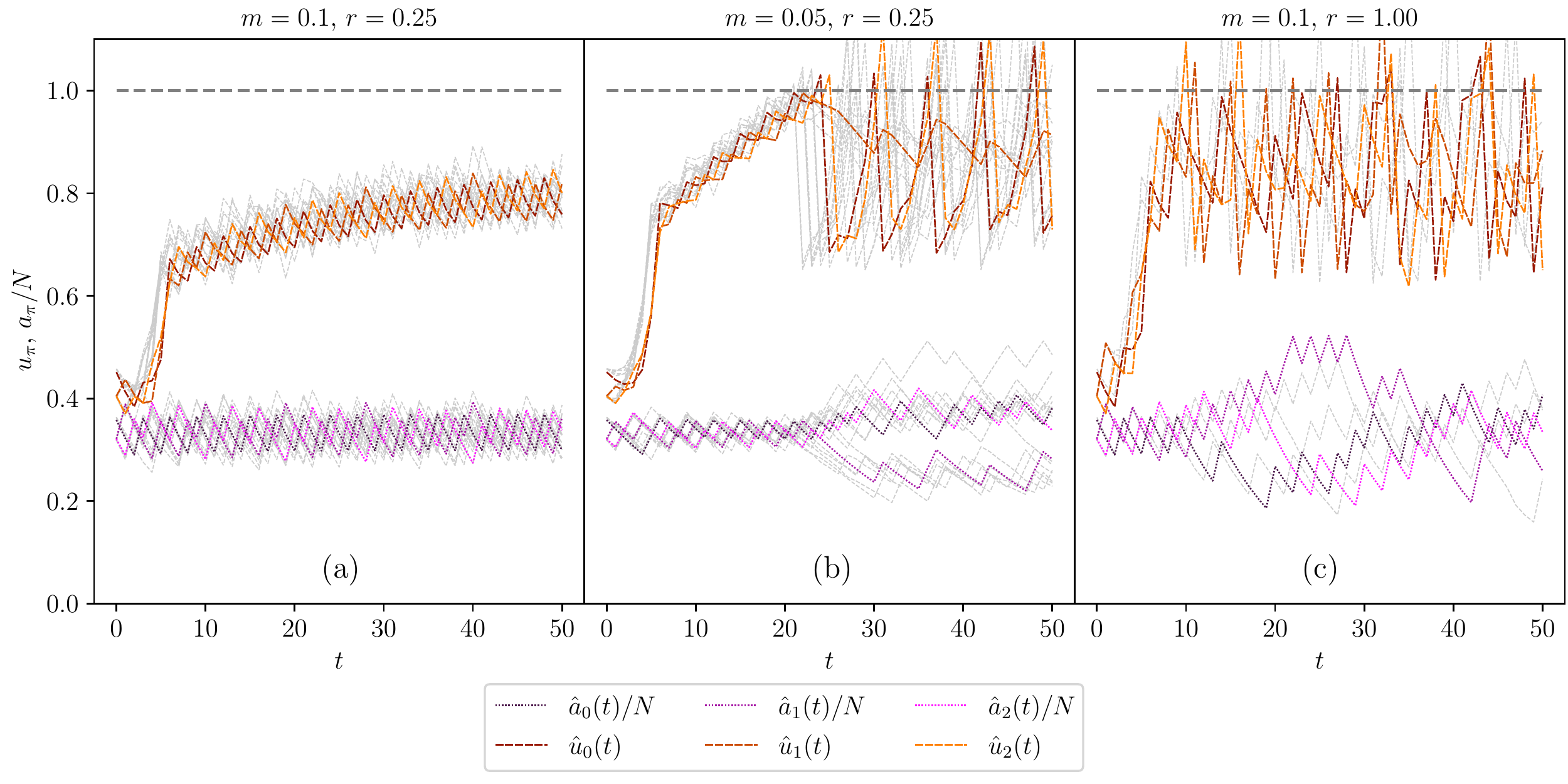}
        \caption{$P=3$, $\alpha_{\mathrm{S}}(\tau) = 2^\tau \text{ if } \tau < 5 \text{ else } 1$.}
        \label{fig:approximation-accuracy:P3:SLOWSTART}
    \end{subfigure}
    \begin{subfigure}{\linewidth}
        \centering
        \includegraphics[width=0.85\linewidth]{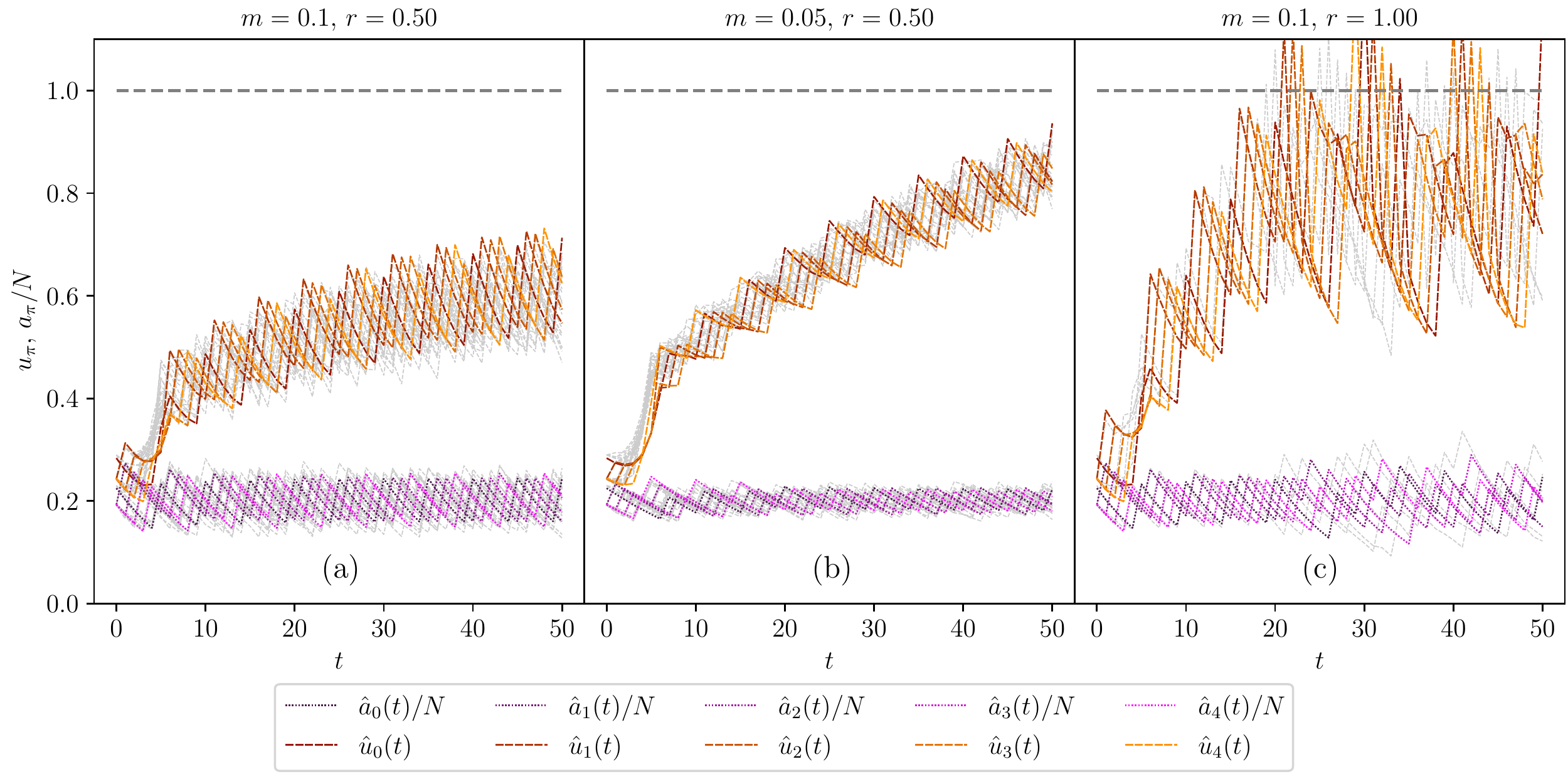}
        \caption{$P=5$, $\alpha_{\mathrm{S}}(\tau) = 2^\tau \text{ if } \tau < 5 \text{ else } 1$.}
        \label{fig:approximation-accuracy:P5:SLOWSTART}
    \end{subfigure}
    \caption{Comparison of model and simulations to
    demonstrate approximation accuracy.}
    \label{fig:approximation-accuracy:appendix}
\end{figure}

The analysis above is repeated for more paths
and a non-constant additive-increase function
in~\cref{sec:appendix:approximation-accuracy}.
In particular,
we repeat this analysis for 
constant additive increase, but with $P =5$ 
(cf.~\cref{fig:approximation-accuracy:P5:CONSTANT}),
as well as with an additive-increase function
that mimics TCP slow-start behavior 
($\alpha_{\mathrm{S}}(\tau) = 2^\tau \text{ if } \tau < 5 \text{ else } 1$)
for both~$P=3$ (cf.~\cref{fig:approximation-accuracy:P3:SLOWSTART})
and~$P=5$ (cf.~\cref{fig:approximation-accuracy:P5:SLOWSTART}).

\section{Logical Consistency of \textit{P}-Step Oscillation}
\label{sec:appendix:p-step-oscillation-consistency}

In~\cref{sec:lossless:flow}, we have shown that
given $P$-step oscillation and without
capacity limits, the flow dynamics exponentially
converge to a dynamic equilibrium where
the rank-$p$ path carries flow volume~$\hat{f}^{(p)}$
in every time step. The general rank-$p$ equilibrium
flow volume is given by the following term:
\begin{equation}
    \begin{split}
    \hat{f}^{(p)} &= \frac{\sum_{p'=0}^{p-1}(1-m)^{p-p'}\hat{\alpha}^{(p')}\cdot \hat{a}^{(p')} + (1-m)^p\cdot\hat{\alpha}^{(P-1)}\cdot \hat{a}^{(P-1)}}{1-\big(1+m\cdot r\cdot z(m,P)\big)\cdot(1-m)^{P-1}}\\
    &+ \frac{ (1+m\cdot r\cdot z(m, P)\big)\cdot\sum_{p'=p}^{P-2} (1-m)^{P-1+p-p'}\cdot\hat{\alpha}^{(p')}\cdot \hat{a}^{(p')}}{1-\big(1+m\cdot r\cdot z(m,P)\big)\cdot(1-m)^{P-1}}
    \end{split}
    \label{eq:lossless:rank-p}
\end{equation}

Interestingly, analyzing
the equilibrium flow volumes~$\{\hat{f}^{(p)}\}_{p\in[P]}$
allows to draw conclusions about the occurrence
of~$P$-step oscillation for a certain
parameter combination, which works by
logical contraposition: If~$P$-step oscillation
occurs for a certain parameter combination,
then~$P$-step oscillation produces
the equilibrium flow volumes~$\{\hat{f}^{(p)}\}_{p\in[P]}$.
However, if the equilibrium flow volumes are themselves 
inconsistent with~$P$-step oscillation, i.e.,
if~$\hat{f}^{(p)} < \hat{f}^{(p+1)}$ for some~$p \in [P-1]$,
then the equilibrium cannot exist and
there is a contradiction. 
This contradiction suggests that~$P$-step oscillation 
is fundamentally impossible 
for the given parameter combination, 
as~$P$-step oscillation would have produced 
the equilibrium flow volumes if it 
had occurred.\footnote{Note that the inverse is not true:
The absence of a contradiction
does not mean that $P$-step
oscillation necessarily occurs for a
given parameter combination.}

Based on this reasoning, we can find a parameter
sub-space for which $P$-step oscillation
is impossible. More precisely,
given any parameter combination,
we can compute the equilibrium flow
volumes~$\{\hat{f}^{(p)}\}_{p\in[P]}$
and check if~$\hat{f}^{(p)} < \hat{f}^{(p+1)}$ 
for any~$p \in [P-1]$. As~\cref{eq:lossless:rank-p}
shows, the parameter space for the
equilibrium flow volumes
consists of the migration rate~$m$,
the reset softness~$r$, the number of
paths~$P$, the additive-increase 
function~$\alpha(\tau)$, and the
number of agents~$N$ (appearing in~$\hat{a}^{(p)}$). 
Luckily, as~$N$ is a linear coefficient
of~$\hat{f}^{(p)}$ and~$N > 0$,
$N$ can be eliminated in the
inequality~$\hat{f}^{(p)} < \hat{f}^{(p+1)}$.
We performed such an exploration of the parameter space
with a focus on~$m$ and~$r$, yielding the results in 
in~\cref{fig:p-step-oscillation-consistency}.
These results indicate that for the
two analyzed additive-increase functions,
$P$-step oscillation is never logically
inconsistent for 2 paths and only
rarely logically inconsistent for
higher number of paths.
While not a definitive proof for the prevalence
of~$P$-step oscillation, these results
suggest that the notion of $P$-step oscillation is
a sound concept for most parameter combinations.

% \section{Variance Calculation with Markov Process}
% \label{sec:appendix:markov-process}

% A single realization
% of the lossless Markov process in~\cref{sec:axioms:equilibria} 
% for~$t$ time steps
% can be represented by a trace~$x \in \{\mathrm{I},\mathrm{M}\}^{t}$,
% where symbol~$I$ corresponds to a time-step
% where agent~$i$ remains on a path and increases its
% congestion-window size, and symbol~M corresponds
% to a time step where agent~$i$ migrates and
% resets its congestion-window size according to~$r$.
% The probability of every trace~$x$ with length~$t$ 
% is~$p_m^{\mathrm{M}(x)}(1-p_m)^{t-\mathrm{M}(x)}$, where~M$(x)$
% yields the number of M-symbols in~$x$.
% Moreover, each trace~$x$ is also associated
% with a congestion-window size, which is given
% by a function~$\Gamma(x)$. For example,
% it holds that~$\Gamma(\mathrm{I}) = \alpha(0)$,
% $\Gamma(\mathrm{II}) = \alpha(0) + \alpha(1)$,
% and~$\Gamma(\mathrm{IMI}) = r\cdot\alpha(0) + \alpha(0)$.
% As a result, the expected congestion-window
% size after~$t$ time steps can be computed
% as follows:
% \begin{equation}
%     \expectation_{i\in A}\big[\mathit{cwnd}_i(t)\big] = \sum\nolimits_{t'=0}^{t} p_m^{t'}\cdot(1-p_m)^{t-t'} \cdot \sum\nolimits_{x \in X(t', t-t')} \Gamma(x)
%     \label{eq:ax:fairness:lossless:formula}
% \end{equation} where~$X(t', t-t')$ is the set of all traces~$x$
% of length~$t$ with~$M(x) = t'$. 
% While this formulation allows to exactly
% calculate the expectation of the congestion-window
% size and its variance, the formula
% becomes computationally intractable for
% high~$t$ due to its combinatorial nature.

\section{Additional Figures}
\label{sec:appendix:additional-figures}

\begin{figure}[H]
\centering
\begin{minipage}{.48\textwidth}
  \centering
  \includegraphics[width=\linewidth,trim=0 0 10 0]{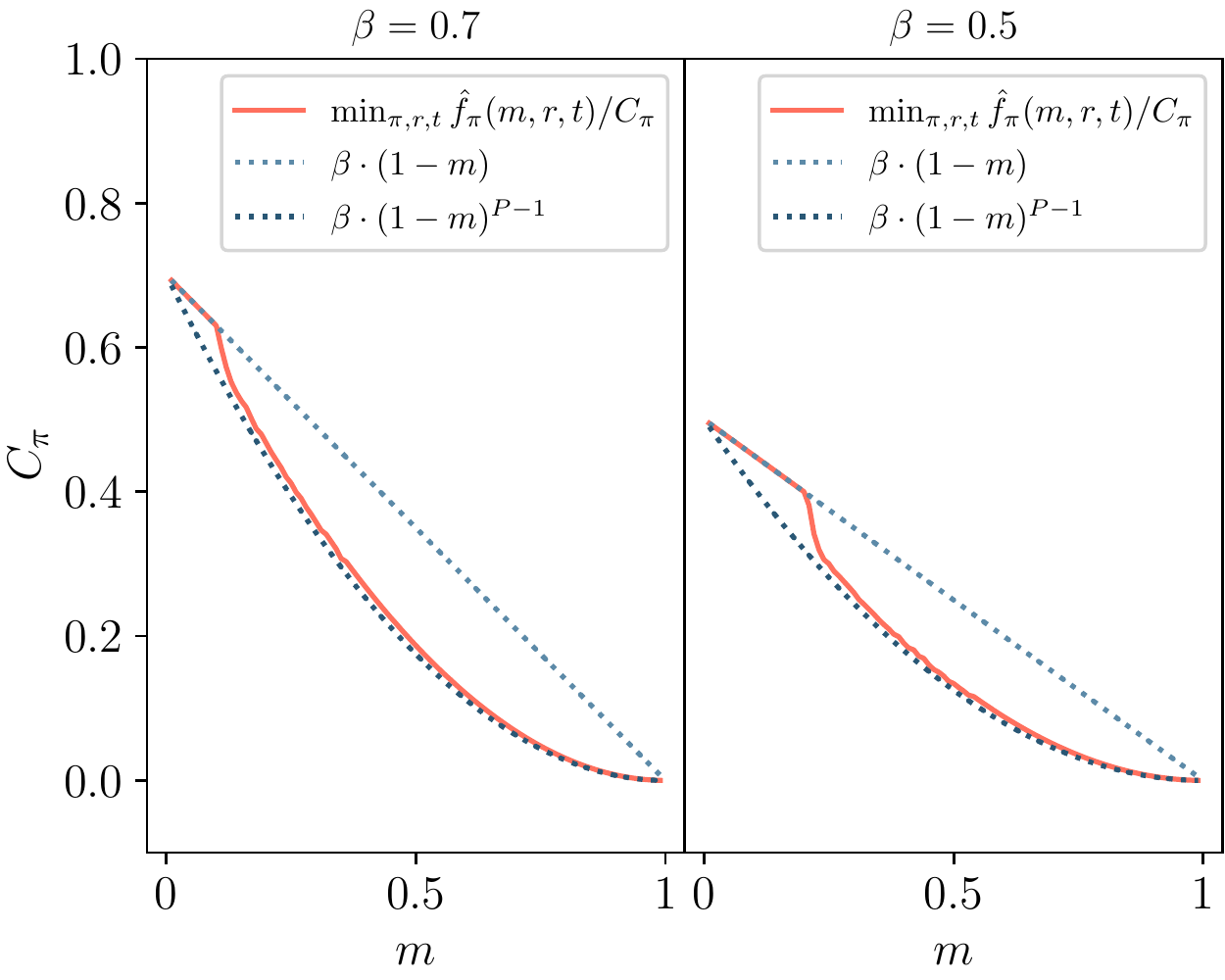}
  \caption{Simulation-based validation of lower bounds on efficiency~$\epsilon$
  for lossy equilibria as derived in~\cref{sec:lossy:flow}. Simulation parameters
  of interest include~$C_{\pi} = 12000$, $N = 1000$, and
  $\alpha(\tau) = 1$.}
  \label{fig:lower-bound-validation}
\end{minipage}
\hfill
\begin{minipage}{.48\textwidth}
    \centering
  \includegraphics[width=\linewidth,trim=10 0 0 0]{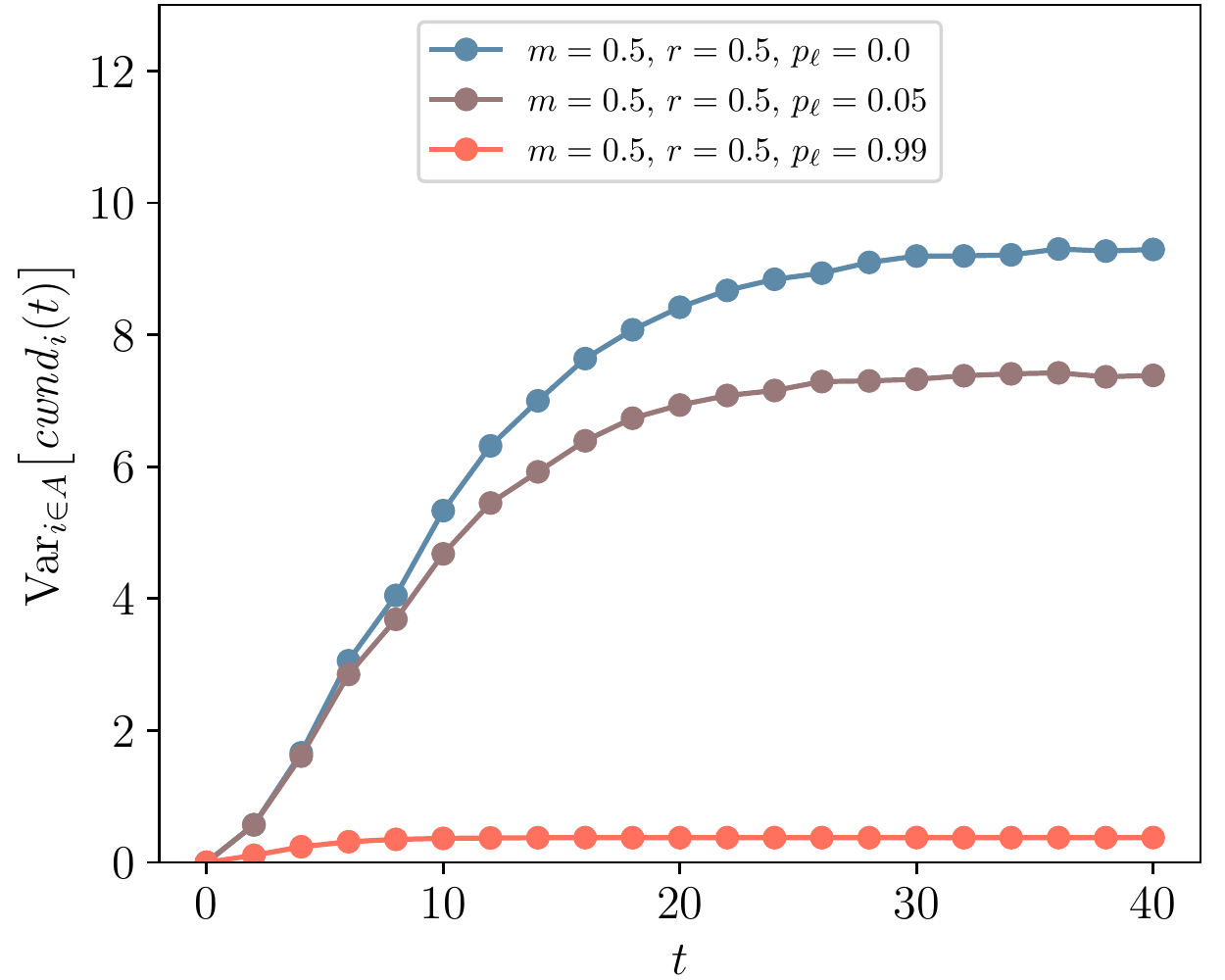}
  \caption{Simulation-based computation of variance in congestion-window size
  according to the lossy Markov process in~\cref{fig:axioms:fairness:lossy:markov}
  for different values of loss probability~$p_{\ell}$.}
  \label{fig:axioms:fairness:variance:lossy}
\end{minipage}
\end{figure}

This appendix section contains
additional figures that illustrate
concepts presented in the main body
of the paper.
\cref{fig:lower-bound-validation} presents
a simulation-based validation of the lower
bounds on the flow volume in lossy equilibria,
derived in~\cref{sec:lossy:flow}.
\cref{fig:axioms:fairness:variance:lossy}
presents the variance in congestion-window
size given a lossy equilibrium, computed
from simulation of the lossy Markov
process 
in~\cref{fig:axioms:fairness:lossy:markov}
in~\cref{sec:axioms:equilibria}.

\end{document}